\documentclass{aa}  
\usepackage{amsmath}	
\usepackage{relsize}
\usepackage{booktabs}
\usepackage{graphicx}
\usepackage{hyperref}
\usepackage{tablefootnote}
\usepackage{gensymb}
\usepackage{xcolor}
\hypersetup{
    colorlinks = true, 
    urlcolor = cyan, 
    linkcolor = blue, 
    citecolor = blue 
}
\usepackage{txfonts}
\usepackage{siunitx}
\usepackage{footnote}
\makesavenoteenv{figure}  
\usepackage{xcolor}
\usepackage{lscape}
\newcommand{\Msun}{\mbox{$\mathrm{M}_\odot$}}

\begin{document} 
   \title{Identification and characterization of optical companions to the population of millisecond pulsars in the globular cluster M3}
   \titlerunning{MSPs in M3 (NGC 5272)}

   \author{Greta Ettorre
          \inst{1,2},
          Emanuele Dalessandro
          \inst{2},
          Mario Cadelano
          \inst{1,2},
          Cristina Pallanca
          \inst{1,2},\\
          Paulo C. C. Freire
          \inst{3}, 
          Alessandro Ridolfi
          \inst{4}
          }

   \authorrunning{Ettorre et al.}         
   \institute{Department of Physics and Astronomy “Augusto Righi”, University of Bologna, Via Gobetti 93/2, 40129 Bologna, Italy\\
              \email{greta.ettorre@inaf.it}
         \and
             INAF – Astrophysics and Space Science Observatory of Bologna, Via Gobetti 93/3, 40129 Bologna, Italy
        \and 
            Max-Planck-Institut f\"ur Radioastronomie, auf dem H\"ugel 69, 53121, Bonn, Germany
        \and
            Faculty of Physics, University of Bielefeld, Universitätsstraße 4, 33615 Bielefeld, Germany
}
   \date{Received XXX; accepted YYY} 
  \abstract
  {The study of binary millisecond pulsars (MSPs) in globular clusters (GCs) is a key ingredient to study binary and stellar evolution under extreme conditions and
it provides interesting insights into the physical properties of their host stellar system.}
  {In this context, an accurate analysis of the optical emission, which is mostly dominated by the companion star, is essential for a comprehensive characterization of these systems and their role within their environment.
  In this work, we present a multi-wavelength investigation of five binary MSPs in the Galactic GC M3 (NGC 5272) using archival Hubble Space Telescope (HST) data. Our analysis builds on phase-connected timing solutions recently obtained with the Five-hundred-meter Aperture Spherical radio Telescope. }
  {Using deep HST images, we investigate the optical counterparts of MSPs M3A, M3B, M3D, M3E, and M3F. For each MSP, we carry out precise astrometric cross-matching with the highly accurate radio positions to identify potential counterparts. When a match is found, we analyze its location in the color-magnitude diagrams and compare the results with updated binary evolution models to infer the system properties.}
  {We confirm the identification of the optical companion to M3B, consistent with the source previously reported in an earlier work, and successfully identify and characterize the optical companions to M3D and M3F. All three are consistent with helium white dwarfs, as expected from the canonical formation scenario. For M3A and M3E, no reliable counterparts are found, but we place strong upper limits on the brightness and mass of the undetected companion. In the case of M3E, we detect a red object near the radio position in two F814W observations; however, astrometric measurements over a 15-year baseline reveal a significant proper motion inconsistent with cluster membership, identifying the source as a foreground contaminant.} {This study highlights the effectiveness of combining precise radio timing with deep, multi-band HST images to uncover and constrain the nature of MSP companions in GCs, offering insights into their formation and evolutionary histories.}


   \maketitle
%
%-------------------------------------------------------------------

\section{Introduction}\label{intro}
Millisecond pulsars (MSPs) are formed in binary systems containing a slowly rotating neutron star (NS) that is eventually spun up to millisecond periods by heavy mass and angular momentum accretion from an evolving companion. At the end of the accretion process, during which these systems are observed as low mass X-ray binaries (LMXBs), these NSs become detectable as radio MSPs. In turn, the companion star is expected to become a white dwarf (WD) with a Helium (He) core \citep{Alpar1982,Stairs2004,Tauris2011,Ferraro2015b}. 

Globular clusters (GCs) provide ideal environments for the formation of these objects. In fact, the dynamical interactions occurring in their very dense cores ($10^{3}-10^{6} \, \rm{pc}^{-3}$) can promote the formation of peculiar stellar populations (i.e., exotic objects), including LMXBs, cataclysmic variables and binaries suitable for recycling NSs into MSPs \citep{Paresce1992,Bailyn1995,Bellanzini1995,Ferraro2001,Pooley2006,Freire2008}. 
Consequently, the number of LMXBs, and therefore MSPs, per unit mass in the Galactic GC population is approximately $10^3$ times higher than in the Galactic field \citep{Clark1975,Hui2010,Turk2013,ZhaoHeinke2022}. Currently, 350 pulsars (PSRs) are known in 46 GCs \footnote{For an up-to-date number visit: \url{https://www3.mpifr-bonn.mpg.de/staff/pfreire/GCpsr.html}}.

Although the standard evolutionary scenario has been widely supported, several exceptions have emerged. These include MSPs with massive CO WD companions \citep{Tauris2011,Pallanca2013b} or double NS systems \citep{Jacoby2006,Tauris2017,Yujie2025}. Moreover, the synergy of multiwavelength observations is found to be crucial in the identification of candidate super-massive pulsars \citep[see for example the M13F system found by][]{Cadelano2020}. Of particular interest are the so-called “holy grail” systems, expected to be formed exclusively in the extreme environments of GC cores through dynamical interactions. One such candidate is NGC~1851E, recently identified by \citet{Barr2024}, which represents the first known binary system hosting a pulsar in orbit around a mass-gap companion, either an exceptionally massive NS or the lowest-mass black hole discovered to date. 

Another class of MSPs deviating from the canonical evolutionary path includes the eclipsing systems with low-mass, non-degenerate companions \citep{Pallanca2010,Roberts2013,Mucciarelli2013,Cadelano2015,Kirichenko2024,Smirnov2025,Turchetta2025}. These systems, collectively referred to as "spiders", are typically subdivided into two main categories: Black Widows (BWs) and Redbacks (RBs).  The distinction between the two types of spiders lies in the mass of their companion stars. RBs have companions with masses ranging from $0.1-0.5$ $\mathrm{M_{\odot}}$, whereas BWs have companions with masses smaller than $0.05$ $\mathrm{M_{\odot}}$. 

These systems are characterized by their small orbital eccentricities, tight orbits (orbital periods P $\lesssim 1$ day), and small mass functions, indicating the presence of a low-mass companion. Moreover, they exhibit periodic eclipses of the radio signal for a significant portion of their orbit, typically during the superior conjunction of the PSR, due to absorption or scattering of radio photons by ionised material between the emitter and the observer \citep{Fruchter1988,Ferraro2001,Ray2012,Cadelano2015,Roberts2013}. 

In this context, multi-wavelength studies of these systems provide a powerful tool for probing their evolution and the physics of accretion processes and for understanding their role within their surrounding environments \citep{Kaluzny2003,Edmonds2002,Pallanca2010,Pallanca2014,Li2014,Cadelano2015,Cadelano2017,Cadelano2020,Turchetta2025,Ettorre2025}. 
In particular, the identification of the optical counterpart to MSPs allows us to explore how the dynamical interactions shape the formation and evolution of MSPs, and to study stellar evolution under extreme conditions  \citep{Ferraro2003,Ferraro2003c,Sabbi2003a,Sabbi2003b,Mucciarelli2013,Ferraro2015b}.

In systems hosting WD companions, the comparison of observed properties with binary evolution models enables estimates of the companion masses and their cooling ages \citep{Ferraro2003b,Cadelano2019,Cadelano2020}. The latter, provide a more robust estimate of the system's true age compared to the characteriztic spin-down ages inferred from radio observations \citep[see][]{Tauris2012}. In particular, precise age determinations are also fundamental for understanding the spin evolution of MSPs and for constraining the physical mechanisms at play during the recycling process \citep{vanKerkwijk2005}.
Finally, when combined with radio timing, optical studies can yield precise mass measurements for both components of the binary system \citep{Ferraro2003b,Cocozza2006,Antoniadis2013,Mucciarelli2013}. These measurements benefit significantly from the well-determined distances and extinction values typical of GCs, leading to smaller uncertainties compared to MSPs in the Galactic field. A very recent example of how the synergy between radio and optical observations enables a full characterization of these systems is presented in  \citet{Lian2025} for the MSP M53A.

In this work we search for the companion stars to five MSPs in the GC M3 (NGC 5272), whose timing solutions were recently updated by \citet{Li2024} using the Five-hundred-meter Aperture Spherical radio Telescope \citep[FAST,][]{Nan2011}.
M3 is an old, bright and extensively studied GC, located at a distance of 10 kpc from the Sun (e.g., \citealt{Carretta2009,Dotter2010,Dalessandro2013}). It is characterized by an intermediate metallicity ($\rm{[Fe/H]} = -1.5$) and negligible reddening \citep[$\rm{E(B-V)} = 0.01$,][]{Harris2010}, given its high galactic latitude \citep[$b=78.71\degree$;][]{Vasiliev&Baumgardt2021}.

Currently, five PSRs are known to be present in M3, all of them in binary systems. PSRs M3A to D were found by \citet{Hessels2007} using the Arecibo telescope, while M3E and M3F were found with FAST \citep{Pan2021}, and their timing solutions were obtained for the first time by \citet{Li2024}. A sixth pulsar candidate, M3C, has not been confirmed. Table~\ref{pulsar_params} and Table~\ref{m3d_m3f_solutions} summarize the main radio timing parameters for the five pulsars in M3, as reported by \citet{Li2024}, along with the physical properties of the binary systems that we were able to characterize, following the methodology described in Sect.~\ref{methods}. The full list of radio timing parameters is available in table~1 of \citet{Li2024}.
\renewcommand{\arraystretch}{1.3}
\begin{table*}[h!]
\setlength{\tabcolsep}{4pt}
\caption{\label{pulsar_params}Main radio timing parameters from \citet{Li2024} and physical properties derived following the methodology outlined in Sect.~\ref{methods}, for M3A, M3B and M3E.}
\centering
\begin{tabular}{lcccc}
\hline\hline
Parameter & M3A & M3B &  \multicolumn{2}{c}{M3E}  \\
\hline
\multicolumn{5}{c}{{Radio timing parameters}}\\
\hline
$\alpha$ (J2000) & 13:42:13.79168(8) & 13:42:11.08686(2) & \multicolumn{2}{c}{13:42:12.3908(4)} \\
$\delta$ (J2000) & 28:23:01.966(1) & 28:22:40.0844(2) & \multicolumn{2}{c}{28:23:16.131(6)} \\
$\mathrm{P_b}$ (days) & 0.1358561044(3) & 1.4173522991(3) & \multicolumn{2}{c}{7.0968527(1)} \\
$\mathrm{P}$ (ms) & 2.54473824519474(6) & 2.38942075782683(5) & \multicolumn{2}{c}{5.4728653539829(3)} \\
$\mathrm{x_p}$ (lt-s) & 0.0256212(7) & 1.8756549(3) & \multicolumn{2}{c}{5.33923(1)} \\
DM ($\mathrm{pc\,cm}^{-3}$) & 26.4278(8) & 26.1510(4) & \multicolumn{2}{c}{26.521(7)} \\
$\textit{e}$ & – & 0.0000040(3) & \multicolumn{2}{c}{0.000309(5)} \\
$\theta_{\perp}$ (arcmin) & 0.6204 & 0.1214 & \multicolumn{2}{c}{0.6545} \\
\hline
\multicolumn{5}{c}{Inferred physical parameters}\\
 & & & WD & MS \\
\hline
$\mathrm{M_{COM}}\,(\Msun)$ & $\leq0.108^{+0.001}_{-0.001}$ & $0.18^{+0.01}_{-0.01}$ & $\geq0.390^{+0.001}_{-0.001}$ & $\leq0.484^{+0.001}_{-0.001}$ \\
$\log{\mathrm{T_{eff}}}$ (K) & – & $4.08^{+0.04}_{-0.04}$ & – & –\\
Cooling age (Gyr) & – & $0.98^{+0.18}_{-0.19}$ & – & –\\
$\mathrm{M_{NS}}\,(\Msun)$ & – & $1.22^{+0.02}_{-0.01}$ & –& – \\
$i$ (deg) & – & $88.84^{+1.16}_{-6.77}$ & –& – \\
\hline
\end{tabular}
\tablefoot{Radio timing parameters: R.A., Dec., orbital period ($\mathrm{P_b}$), spin period ($\mathrm{P}$), projected semi-major axis ($\mathrm{x_p}$), dispersion measure (DM), orbital eccentricity (\textit{e}) and angular offset from the GC center ($\theta_{\perp}$). Inferred parameters: companion mass, effective temperature, WD cooling age, neutron star mass, and orbital inclination. For M3E, the companion mass limit is given under both the assumptions of a WD and a non-degenerate main sequence (MS) companion.}
\end{table*}

\renewcommand{\arraystretch}{1.3}
\begin{table*}[h!]
\setlength{\tabcolsep}{5pt}
\caption{\label{m3d_m3f_solutions}Main radio timing parameters and inferred physical parameters for the M3D and M3F systems, showing both the young and old solutions (see Sections~\ref{M3D} and \ref{M3F}).}
\centering
\begin{tabular}{lcccc}
\hline\hline
 Parameter & \multicolumn{2}{c}{M3D} & \multicolumn{2}{c}{M3F} \\
\hline
\multicolumn{5}{c}{Radio timing parameters}\\
\hline
$\alpha$ (J2000) & \multicolumn{2}{c}{13:42:11.35220(9)} & \multicolumn{2}{c}{13:42:12.37908(6)} \\
$\delta$ (J2000) & \multicolumn{2}{c}{28:22:30.160(1)} & \multicolumn{2}{c}{28:22:37.270(1)} \\
$\mathrm{P_b}$ (days) & \multicolumn{2}{c}{128.7454866(5)} & \multicolumn{2}{c}{2.991987692(4)} \\
$\mathrm{P}$ (ms) & \multicolumn{2}{c}{5.4429751130405(3)} & \multicolumn{2}{c}{4.4038055250797(2)} \\
$\mathrm{x_p}$ (lt-s) & \multicolumn{2}{c}{38.521421(2)} & \multicolumn{2}{c}{2.225893(4)} \\
DM ($\mathrm{pc\,cm}^{-3}$) & \multicolumn{2}{c}{26.3677(3)} & \multicolumn{2}{c}{26.436(2)} \\
$\textit{e}$ & \multicolumn{2}{c}{$7.47556(1)\times10^{-2}$} & \multicolumn{2}{c}{–} \\
$\theta_{\perp}$ (arcmin) & \multicolumn{2}{c}{0.1464} & \multicolumn{2}{c}{0.1677} \\
\hline
\multicolumn{5}{c}{Inferred physical parameters}\\
\cmidrule(lr){2-3} \cmidrule(lr){4-5}
 & Young & Old & Young & Old \\
\hline
$\mathrm{M_{COM}}\,(\Msun)$ & $0.33^{+0.06}_{-0.07}$ & $0.20^{+0.02}_{-0.02}$ & $0.32^{+0.07}_{-0.06}$ & $0.21^{+0.02}_{-0.02}$ \\
$\log{\mathrm{T_{eff}}}$ (K) & $4.14^{+0.05}_{-0.06}$ & $4.00^{+0.04}_{-0.03}$ & $4.07^{+0.06}_{-0.04}$ & $3.98^{+0.02}_{-0.02}$ \\
Cooling age (Gyr) & $0.31^{+0.15}_{-0.11}$ & $3.93^{+1.19}_{-1.18}$ & $0.51^{+0.29}_{-0.18}$ & $5.15^{+1.20}_{-1.19}$ \\
$\mathrm{M_{NS}}\,(\Msun)$ & $1.72^{+0.24}_{-0.32}$ & $1.28^{+0.09}_{-0.05}$ & $1.70^{+0.25}_{-0.32}$ & $1.72^{+0.22}_{-0.31}$ \\
$i$ (deg) & $47.33^{+7.91}_{-7.13}$ & $81.43^{+5.89}_{-7.88}$ & $32.46^{+4.61}_{-4.34}$ & $54.08^{+7.06}_{-8.35}$ \\
\hline
\end{tabular}
\tablefoot{For each system, two solutions are shown: the young and the old ones, as identified with cyan and magenta colors, respectively, in Fig.~\ref{solutionM3D} and Fig.~\ref{solutionM3F}.}
\end{table*}
The companion star of M3B was identified by \citet{Cadelano2019} as an extremely low-mass WD with a mass of only $0.19\,\rm{M}_{\odot}$. They also proposed a He WD as the likely counterpart to M3D. However, for this system, they adopted the source coordinates from the timing solution of \citet{Hessels2007}, which are affected by significantly larger uncertainties ($0\arcsec\!.6$ in R.A. and a much larger uncertainty of $14\arcsec$ in Dec.), compared to the more recent solution by \citet{Li2024}, possibly hindering the identification of the true companion.

Building on the recent and precise phase-connected timing solutions provided by \citet{Li2024} for all the confirmed pulsars in M3, we create an extended HST dataset, spanning from the near-UV to the optical I band, to search for their optical companions and infer their properties using updated stellar models. While the optical companion to M3B was previously characterized by \citet{Cadelano2019}, we revisit this system using new photometry and improved models to derive its parameters. For the remaining MSPs, this work presents the first systematic search and, where possible, characterization of their optical counterparts.
The paper is organised as follows: In Sect.~\ref{analysis}, we present the HST dataset and describe the photometric analysis, including the calibration and astrometric procedures adopted during the data reduction. Sect.~\ref{methods} outlines our approach for identifying the optical counterparts and deriving their physical properties, and provides a detailed discussion of the results for the five MSPs. Finally, Sect.~\ref{Conclusions} summarizes our main findings and conclusions.

\section{HST observations and data reduction}\label{analysis}
\begin{figure}
    \centering
    \includegraphics[width=0.95\hsize]{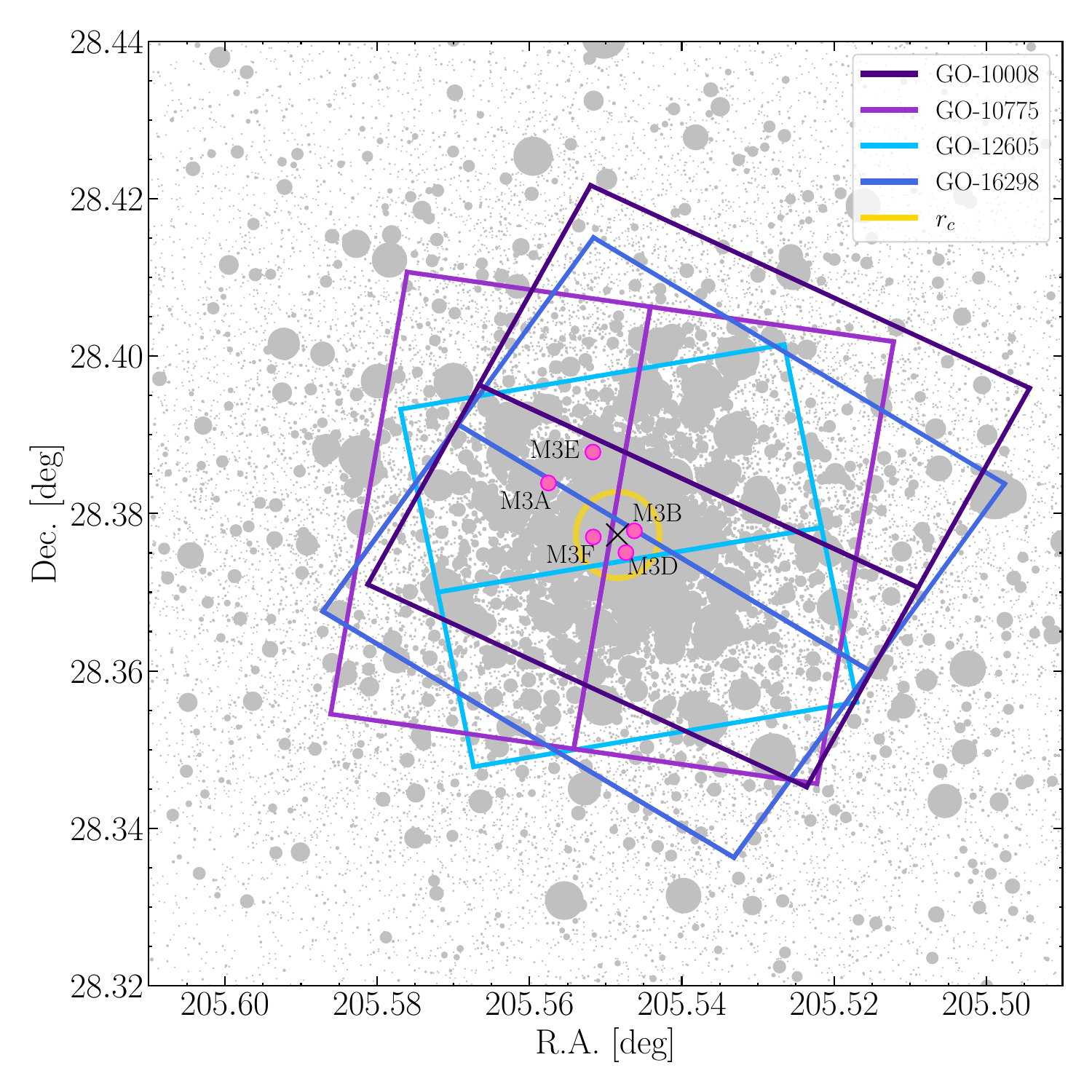}
    \caption[]{Footprints of the FoVs covered by the HST observations listed in Table~\ref{dataset}. The underlying stars come from the Gaia DR3 dataset \citep{Gaia2023} centerd on M3, obtained using the Gaia \textit{G}-band magnitude to scale the size of the data points. Each FoV is color-coded according to the GO proposal number, listed in Table~\ref{dataset}. The pink dots represent the position of each MSP and their name is represented. The black cross marks the center of the cluster (R.A. $=205\degree\!.548416$, Dec. $=28\degree\!.377277$) and the yellow circle is the core radius $\mathrm{r_c}=0.98$ pc. Both values are taken from the 2023 version of the GC catalog provided by \citet[][but see also \cite{Vasiliev&Baumgardt2021}]{Baumgardt2020}\protect\footnotemark.}
    \label{map}
\end{figure}

In this work, we employed an extended photometric dataset consisting of high-resolution HST images ranging from the near-UV (F225W) to the optical I band (F814W), obtained using the ultraviolet-visible (UVIS) channel of the Wide Field Camera 3 (WFC3) and  Wide Field Channel (WFC) of the Advanced Camera for Surveys (ACS). Table \ref{dataset} provides an overview of the images used in this work to build our photometric catalog, while Fig. \ref{map} shows the footprints of the adopted HST observations. 
\footnotetext{\url{https://people.smp.uq.edu.au/HolgerBaumgardt/globular/}}

The WFC3/UVIS dataset comprises 14 images acquired on 2012 May 15 with the F275W, F336W, and F438W filters, as part of proposal 12605 (PI: Piotto). The ACS/WFC data consist of observations taken across three distinct epochs. The first epoch includes 4 images obtained on 2004 May 9 in the F435W and F555W filters (proposal 10008, PI: Grindlay). The second set comprises 10 images taken on 2006 February 20 with the F606W and F814W filters (proposal 10755, PI: Sarajedini). Finally, the third ACS/WFC dataset, from proposal 16298 (PI: Libralato), consists of 5 images in the F814W filter acquired on 2021 June 3. In total, we used 33 images spanning seven different filters and four observational epochs. 

Our photometric reduction process is tailored to achieve high accuracy and completeness at faint magnitudes in crowded fields, an essential requirement for identifying the optical counterparts of MSPs within GC cores. The photometric analysis was performed independently for each of the four dataset, following the general procedure outlined below.

In detail, the data reduction was carried out on the calibrated exposures that include the charge transfer efficiency (CTE) correction (i.e., \texttt{$\_$flc} images). For the F275W, F336W, and F438W exposures, cosmic ray removal was performed using the \texttt{lacosmic} Python package \citep{VanDokkum2001}. After correcting for the `Pixel-Area Map (PAM)', the photometric analysis was conducted using the \texttt{DAOPHOT} package \citep{Stetson1987}. As a first step, we generated a source list using the \texttt{FIND} routine, selecting only those detections with peak counts exceeding a specified threshold above the local background noise. The detection threshold was set empirically for each dataset to effectively minimize contamination from background fluctuations and spurious detections. Specifically we chose a $3\sigma$ threshold for the WFC3/UVIS dataset and the ACS/WFC dataset from proposals 16298 and 10755, while $5\sigma$ for the F435W and F555W images, where $\sigma$ denotes the standard deviation of the local background.  Using \texttt{PHOTO} we obtained the concentric aperture photometry of each source. For each image we selected a large number ($\sim 200$) of bright, unsaturated and isolated stars. 

These stars were subsequently used to model the point-spread function (PSF) using the \texttt{PSF} routine. The resulting best-fit PSF models were then applied to all detected sources with \texttt{ALLSTAR}. To refine the photometry, we employed the \texttt{ALLFRAME} package \citep{Stetson1994}, which enables simultaneous PSF-fitting across all available images within each dataset. This step required the construction of a master list containing all objects to be fitted. We used \texttt{DAOMATCH} and \texttt{DAOMASTER} to align the catalogs and produce, for each filter, a list of stars with their average magnitudes and associated uncertainties, computed from the dispersion of individual measurements. Only stars detected in at least half of the exposures +1 were retained. The resulting single-filter lists were then merged into a comprehensive master list using again \texttt{DAOMATCH} and \texttt{DAOMASTER}.

Both the WFC3/UVIS and the ACS/WFC images are characterized by significant geometrical distortions across the FoV. Therefore, we corrected the instrumental positions of the WFC3 dataset with the equations by \cite{Bellini2009} for the filter F275W, while we used equations by \citet{meurer2004} for the ACS/WFC images. 

We assigned a photometric quality flag to each star in the catalog by selecting sources based on the \texttt{CHI} and \texttt{SHARP} photometric parameters, following the procedure described in \citet{Ettorre2025}.

For the WFC3/UVIS dataset, we calibrated the instrumental magnitudes to the VEGAMAG photometric system by cross-matching our catalog with the HST UV Globular Cluster Survey (HUGS) catalog \citep{Piotto2015, Nardiello2018}, using the \texttt{CataXcorr} software \citep{Montegriffo1995}. We selected only stars with good photometric quality. We then computed the average magnitude difference between stars in common to both catalogs and adopted this value as a zeropoint correction to convert our instrumental magnitudes to the VEGAMAG system.
The same procedure was applied to images from HST proposal ID 10775, using the ACS Survey of Galactic Globular Clusters catalog \citep{Sarajedini2007} as the photometric reference. For all remaining ACS/WFC images, we applied the appropriate transformation equations and zeropoints provided by \citet{Sirianni2005} to calibrate the instrumental magnitudes.

Using \texttt{CataXcorr} we transformed the instrumental coordinates to the absolute coordinate systems $(\alpha,\delta)$ by cross-matching each catalog with the Gaia DR3 dataset \citep{Gaia2023}. The coordinate system used in the Gaia dataset is based on the International Celestial Reference System, enabling accurate comparison with MSP positions obtained from timing analyzes that employ solar system ephemerides, which are also referenced to the same celestial framework. 

The catalogs derived from the photometric analysis of the four datasets were cross-matched and combined, producing the final catalog. This comprehensive catalog includes all stars detected in at least one filter and contains over 350000 sources.

Finally, we took advantage of the long temporal baseline offered by the available multi-epoch dataset (see Table~\ref{dataset}) to perform a relative proper motion (PM) analysis. For this purpose, we selected F814W images from HST programs 10775 (PI: Sarajedini) and 16298 (PI: Libralato), which are separated by approximately 15 years. This combination was chosen as it maximizes the number of stars, particularly the candidate companions to the five MSPs, for which PMs can be reliably measured. 
For each star detected in both F814W datasets, we determined its position at each epoch by averaging the coordinates from individual exposures. The uncertainties were estimated from the standard deviation of these measurements. We then calculated the displacements in the x and y directions as the difference between the mean positions at the two epochs. Dividing these displacements by the time baseline yielded the proper motions in pixels per year. Finally, we converted these values into milliarcseconds per year using the ACS/WFC pixel scale, thereby obtaining the relative PMs.
\begin{table*}[h]
\renewcommand{\arraystretch}{1}  
\caption{\label{dataset}HST observations of M3 used in this work.}
\centering
\begin{tabular}{lcccccc}
\hline\hline
Instrument & Date & Filter & N of images & Proposal ID & PI & $\mathrm{N} \times \mathrm{Exp.\, time \, (s)}$\\
\hline
ACS/WFC   &   2004 May 09  &F435W & 2   & 10008 & Grindlay & $2 \times 340$\\
ACS/WFC   &   2004 May 09  &F555W & 2   & 10008 & Grindlay & $2 \times 339$\\
ACS/WFC   &   2006 Feb 20  &F606W & 5   & 10775 & Sarajedini & $4 \times 130 +12$ \\
ACS/WFC   &   2006 Feb 20  &F814W & 5   & 10775 & Sarajedini & $4 \times 150 +12$\\
WFC3/UVIS   &   2012 May 15 &F275W & 6   & 12605 & Piotto & $6 \times 415$\\
WFC3/UVIS    &   2012 May 15   &F336W & 4  & 12605  & Piotto & $ 4 \times 350$\\
WFC3/UVIS    &   2012 May 15  &F438W  & 4  & 12605 & Piotto & $4 \times 42$\\
ACS/WFC   &   2021 Jun 03  &F814W & 5   & 16298 & Libralato & $4 \times 337 +35$\\
\hline
\end{tabular}
\tablefoot{From left to right: HST instrument name, date of observation, filter of the observation, number of images, proposal ID, principal investigator and exposure time in seconds for each image.}
\end{table*}

\section{Searching and characterizing the optical counterparts}\label{methods}
As it can be noticed from Table~\ref{pulsar_params} and Table~\ref{m3d_m3f_solutions}, the positions of the five MSPs obtained from radio timing analysis have extremely small uncertainties, with a maximum combined uncertainty of $\sim 0.\!\arcsec009$ for M3E. Consequently, the dominant source of positional uncertainty likely arises from our astrometric calibration, specifically from the residuals associated with the polynomial transformations used to align our instrumental positions with the Gaia reference frame. 

Therefore, for each binary MSPs in the cluster, we searched for optical counterparts by carefully inspecting all stars located within a circle of radius $0.\!\arcsec1$ centered on the radio position. For each candidate star, we analyzed its location in the color–magnitude diagrams (CMDs) and investigated possible photometric variability that could be associated with the binary orbital period. 

\begin{figure*}
    \centering
    \includegraphics[width=0.9\hsize]{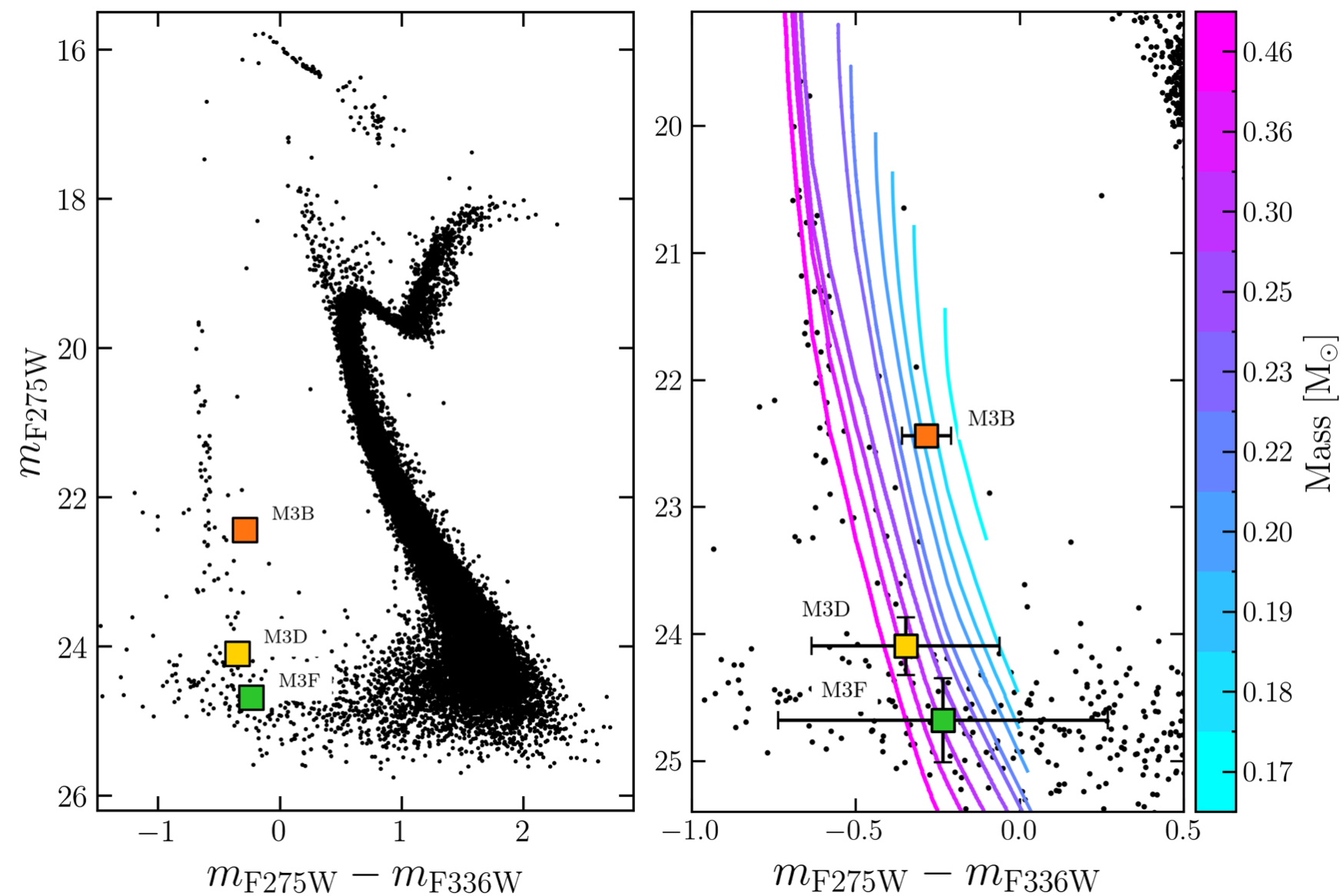}
    \caption{color-magnitude diagram in the ($m_{\rm F275W}$, $m_{\rm F275W} - m_{\rm F336W}$) plane. Black points represent stars from our catalog with good photometric quality. The optical counterparts to M3B, M3D, and M3F are marked with orange, yellow and green squares, respectively. Right panel: Same CMD as shown on the left, but zoomed in to focus on the regions around the counterparts. Error bars for each counterpart reflect the corresponding photometric uncertainties. Overplotted are the He WDs cooling tracks from \citet{Cadelano2020}. These correspond to stellar masses, ordered from right to left, between $0.17\,\Msun$ and $0.46\,\Msun$}, as indicated by the colorbar
    \label{CMD}
\end{figure*}

In the following sections, we present a detailed description of the results from our search and analysis of the five MSPs found in M3. We successfully identified and characterized the optical companions of three MSPs, M3B, M3D, and M3F, as described in details in Sect.~\ref{detections}. In all three cases, the optical source coincides precisely with the pulsar's radio position, demonstrating excellent agreement between the radio timing positions and the astrometry of our catalog, and confirming its high astrometric accuracy.
These counterparts align with the WD cooling sequence in the ($m_{\rm F275W}$, $m_{\rm F275W} - m_{\rm F336W}$) CMD shown in Fig.~\ref{CMD}. For these three counterparts, it was not possible to derive PMs, as they are only detected in WFC3/UVIS images, acquired at the same epoch. 
On the contrary, no plausible optical counterparts were found for M3A and M3E, as detailed in Sect.~\ref{nondetections}.
While the photometric uncertainties of the three identified optical companions, M3D and M3F in particular, make them compatible with being either He or CO WDs, in the following analysis (Sect.~\ref{detections}) we limit our study to He WD models as the vast majority of binary MSPs are expected to form through the canonical evolution scenario, which involves a phase of stable mass-transfer from a Roche-Lobe filling red giant secondary, eventually producing a He-WD companion. 
We stress here that this choice is supported by the  observed properties of both M3D and M3F (see Table~\ref{m3d_m3f_solutions}). In detail, both objects are relatively fast-spinning, with spin periods of  $\sim 5.4$ and 4.4 ms, respectively. Such short spin periods are expected for MSPs with He WD companions, which typically evolve through a phase with stable Roche-lobe overflow over long timescales. This allows the NS to efficiently accrete material and be spun up to short periods. 
In contrast, MSPs with CO WDs generally originate from intermediate-mass X-ray binaries, in which the secondary star, in a post-red giant branch phase (likely an asymptotic giant branch star), transfers mass over a brief episode, leading to slower spin periods \citep[>10 ms;][]{Tauris2012CO}.
In addition, in the case of M3D, the long orbital period of $\sim 129$ days further supports a He WD companion, as CO WD MSPs are expected to have shorter orbital periods (<40 days) \citep{Tauris2012CO}.

\subsection{Detected companions}\label{detections}
\subsubsection{M3B}\label{M3B}
The radio timing study performed by \citet{Li2024} suggests that M3B has a minimum companion mass of $0.21\, \Msun$ and an orbital period of 34.0 hours. As mentioned in the Introduction, the optical companion to this MSP was previously identified and characterized by \citet{Cadelano2019} as an extremely low-mass He WD with a mass of $\mathrm{M_{COM}} = 0.19 \pm 0.02\,\Msun$, a cooling age of $1.0^{+0.2}_{-0.3}$ Gyr, and an effective temperature of $(12 \pm 1) \times 10^3$ K. 

\begin{figure*}
    \centering
    \includegraphics[width=0.9\hsize]{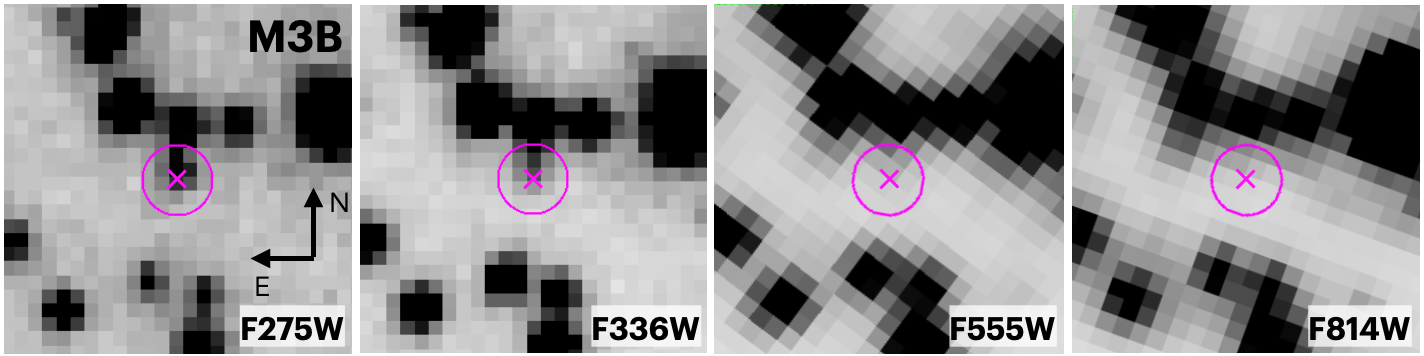}
    \caption{Finding charts for the MSP M3B. The panels display \texttt{\_drc} images in the following HST filters, from left to right: WFC3/UVIS F275W and F336W (proposal ID 12605), ACS/WFC F555W (proposal ID 10008), and F814W (proposal ID 10775).  Each panel shows a $1\arcsec \times 1\arcsec$ field of view centered on the radio position of the pulsar. The magenta cross marks the nominal position of the MSP, while the magenta circle (radius $0.\!\arcsec1$) indicates the search radius. North is up, and East is to the left.}
    \label{chartM3B}
\end{figure*}

In our search for the optical counterpart using our photometric catalog, we confirm the identification of the same source: a very blue object located precisely at the pulsar's radio position, as illustrated in the finding chart in Fig.~\ref{chartM3B}, and consistent with the WD cooling sequence in the CMD (see Fig.~\ref{CMD}). 
We were able to measure its magnitude in the F275W, F336W, and F438W filters, which were subsequently used for comparison with theoretical models. No evidence of photometric variability was found for this object. Although the lack of variability could stem from the limited and uneven coverage of the orbital period in the dataset, it is worth mentioning that such variability is uncommon for degenerate companions. Moreover, in cases where it is observed, it is generally not driven by irradiation from the MSP, but rather by intrinsic processes within the WD, such as global stellar pulsations \citep{Cocozza2006,Maxted2013,Kilic2015}.

At this point we derive the physical parameters of the promising optical counterpart candidate by comparing its magnitudes with He WD stellar evolution models.
The models adopted here are the updated version of the models presented in \citet{Cadelano2019} and already used by \citet{Cadelano2020}, covering a mass range between $0.17$ and $0.47\,\Msun$. While we recall the reader to that paper for a full description of the latter models, here we briefly report the main ingredients.  Binary evolutionary tracks were computed using MESA \citep[version 12115;][]{Paxton2011,Paxton2013,Paxton2015,Paxton2018}, following a procedure similar to that of \citet{Istrate2014,Istrate2016}. The initial binary configuration assumes a $1.4\,\Msun$ NS, treated as a point mass, and a $1.1\,\Msun$ donor star. 
All the available models were interpolated to create a fine grid in the mass range $0.17$–$0.46\,\Msun$, effective temperature between $\sim3000$ and $130.000$ K, and cooling ages extending up to the cluster age.

To quantitatively constrain the companion's properties, such as mass, age, and effective temperature, we defined a multivariate Gaussian logarithmic likelihood. The general equation in the $d$-dimensional case is:
\begin{equation}
\ln \mathcal{L}
= -\frac{1}{2} \Big[ \Delta\mathbf{y}^\top \, \Sigma^{-1} \, \Delta\mathbf{y}
+ \ln |\Sigma| + d \, \ln (2\pi) \Big]
\end{equation}

where $\Delta\mathbf{y}$ is the vector of residuals between the observed and model quantities, defined in the $(m_{275}, m_{275}-m_{336}, m_{336}-m_{438})$ space for M3B, and in the $(m_{275}, m_{275}-m_{336})$ space for M3D and M3F. The matrix $\boldsymbol{\Sigma}$ is the corresponding covariance matrix including photometric, distance modulus, and reddening uncertainties.
This method allowed us to derive constraints on the key parameters of the companion. For each WD parameter, we determined the best estimates and associated uncertainties from the 0.16, 0.50, and 0.84 percentiles of the corresponding distributions.
The values obtained for M3B, listed in Table~\ref{pulsar_params}, are consistent with those previously reported by \citet{Cadelano2019}, within the associated uncertainties.

\begin{figure*}
    \centering
    \includegraphics[width=\hsize]{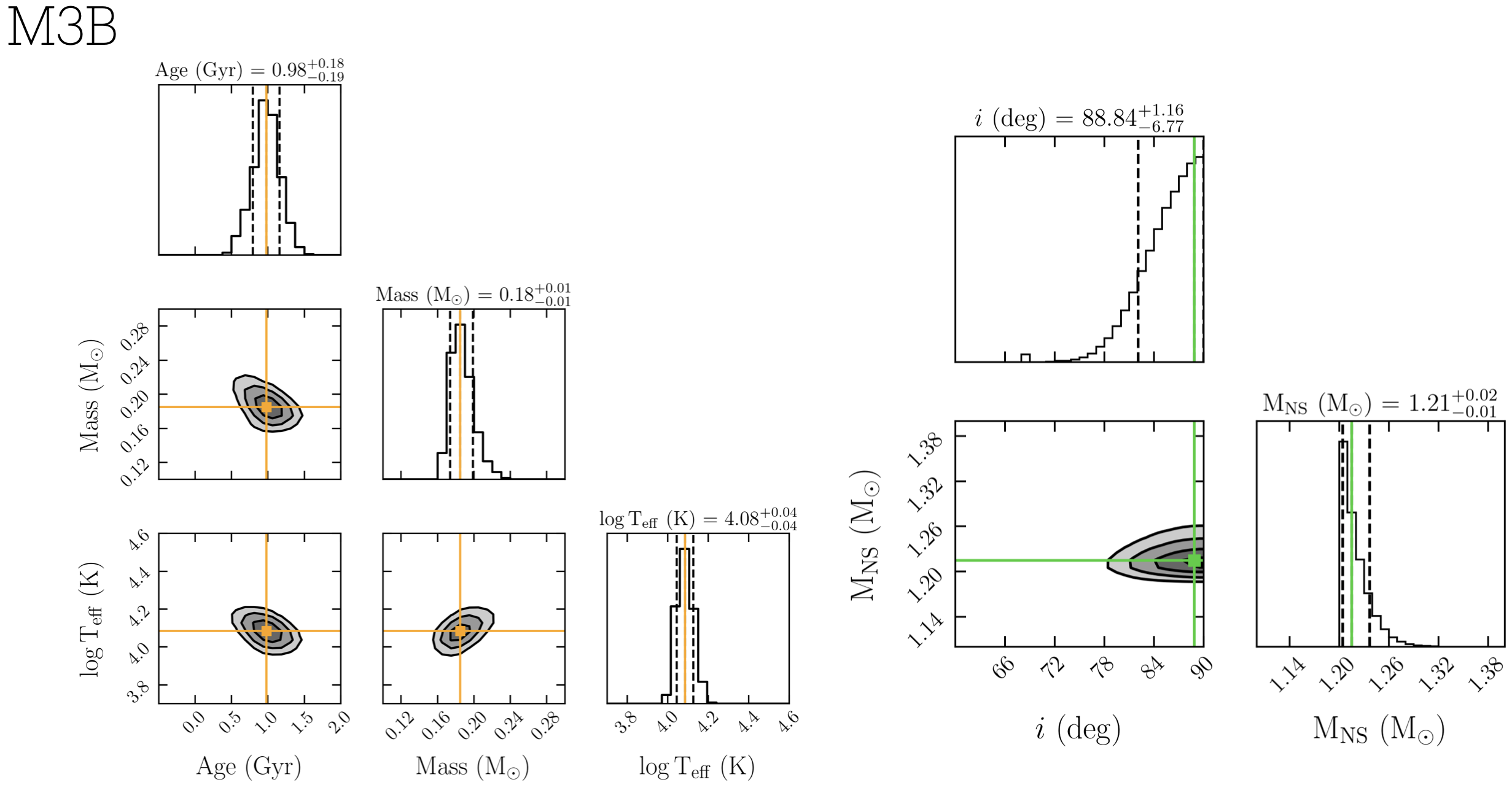}
    \caption{Left-hand corner plots: Constraints on the mass, cooling age, and effective temperature of the companion star to M3B are presented. The 1D histograms show the marginalised likelihood distributions for each parameter, with the solid orange and dashed black lines indicating the 50th, 16th, and 84th percentiles, respectively. These percentiles represent the best-fit values and associated uncertainties. In the 2D histograms, contours correspond to the 1$\sigma$, 2$\sigma$, and 3$\sigma$ levels,  with the best-fit values marked by orange points and lines. The derived values for mass, cooling age, and temperature are reported at the top of each 1D distribution panel. Right-hand corner plot: Constraints on the NS mass and the orbital inclination angle of M3B. The one-dimensional histograms represent the marginalised probability distributions for these two parameters, with the solid green and dashed black lines indicating the best-fit values and their associated uncertainties. The bottom-left panel shows the joint two-dimensional posterior probability distribution, with contours representing the 1$\sigma$, 2$\sigma$, and 3$\sigma$ confidence levels.}
    \label{solutionM3B}
\end{figure*}

The likelihood distributions of the companion's parameters for M3B are shown in the left-hand panel of Fig.~\ref{solutionM3B}. The age distribution peaks at $0.98$ Gyr, and the mass distribution is narrow and symmetric around $\mathrm{M_{COM}} = 0.18\,\Msun$, indicating that this parameter is well constrained and confirming the extremely low mass of the MSP’s companion star. The $\log{\mathrm{T_{eff}}}$ distribution is also symmetric, with a clearly defined peak around $\log{\mathrm{T_{eff}}} =4.08 $ K. 

By determining the companion mass and using the binary orbital parameters derived through radio timing analysis, we can also place constraints on both the NS mass and the orbital inclination of the binary system. This is possible because the masses of the binary components are related to the orbital parameters via the NS mass function:
\begin{equation}\label{massfunction}
    \frac{(\mathrm{M_{COM}}\sin{i})^3}{(\mathrm{M_{COM}}+\mathrm{M_{NS}})^2} = \frac{4\pi^2\mathrm{x_p}^3}{\mathrm{G}\mathrm{P_{ORB}}^2}
\end{equation}
Here, $\mathrm{M_{COM}}$ and $\mathrm{M_{NS}}$ are the masses of the companion star and the NS, respectively, $i$ is the orbital inclination angle, $\mathrm{G}$ the gravitational constant, $\mathrm{x_p}$ is the projected semimajor axis, and $\mathrm{P_{ORB}}$ is the orbital period.
One should notice that the right-hand side of the equation depends only on the binary orbital parameters derived from radio data analysis, making its value well constrained. In contrast, the left-hand side involves the measured companion mass alongside two completely unknown quantities: the NS mass and the orbital inclination angle. 

In order to explore the combination of NS masses and orbital inclination angles able to reproduce the observed mass function, we employed the affine-invariant Markov Chain Monte Carlo (MCMC) ensemble sampler, \texttt{emcee}\footnote{\url{https://emcee.readthedocs.io/en/stable/}} \citep{Foreman-Mackey2013}. We used a uniform prior for the distribution of $\cos{i}$ between 0 and 1, and a flat prior for $\mathrm{M_{NS}}$, within the range of $1.2-2.0\,\Msun$, consistent with the NS mass distribution found by \citet{Abbott2023}. Additionally, we tested a bimodal Gaussian prior on $\mathrm{M_{NS}}$ in the range $1.0$–$2.5\,\Msun$, as suggested by \citet{Antoniadis2016}, with the corresponding results presented in Appendix~\ref{appendix1}.

The best-fit value was determined as the 50th percentile of the inferred distribution, while the lower and upper uncertainties were defined as the 16th and 84th percentiles, respectively.
A similar method was used by \citet{Cadelano2019} to constrain the NS mass and orbital inclination of M3B, yielding $\mathrm{M_{NS}}=1.1^{+0.3}_{-0.3}$ and $i=89^{+1}_{-26}$. We show the updated posterior distributions shown in the right-hand panel of Fig.~\ref{solutionM3B}. For the NS mass we find $\mathrm{M_{NS}}=1.22^{+0.02}_{-0.01}\,\Msun$, a slightly larger value than previously reported in \citet{Cadelano2019}, but consistent within the uncertainties. The inclination distribution retains the truncated Gaussian shape described by \citet{Cadelano2019}, clearly indicating an almost edge-on orientation. As such, we adopt the peak of the distribution as the best estimate and define the lower uncertainty as the 16th percentile of a symmetrised distribution, yielding $i = 88.84^{+1.16}_{-6.77}$ deg.
\subsubsection{M3D}\label{M3D}
This binary system appears to have experienced an unusual evolutionary path, as suggested by the 129-day orbital period derived from radio timing analysis by \citet{Li2024}. Using the position derived from the timing solution by \citet{Hessels2007}, \citet{Cadelano2019} proposed a He WD as the most plausible companion to M3D. However, the radio position of the system was previously affected by significant uncertainties. The improved timing analysis by \citet{Li2024} markedly enhanced the precision of the system’s coordinates, considerably reducing the errors in both R.A. and Dec. relative to earlier measurements. 

\begin{figure*}
    \centering
    \includegraphics[width=0.9\hsize]{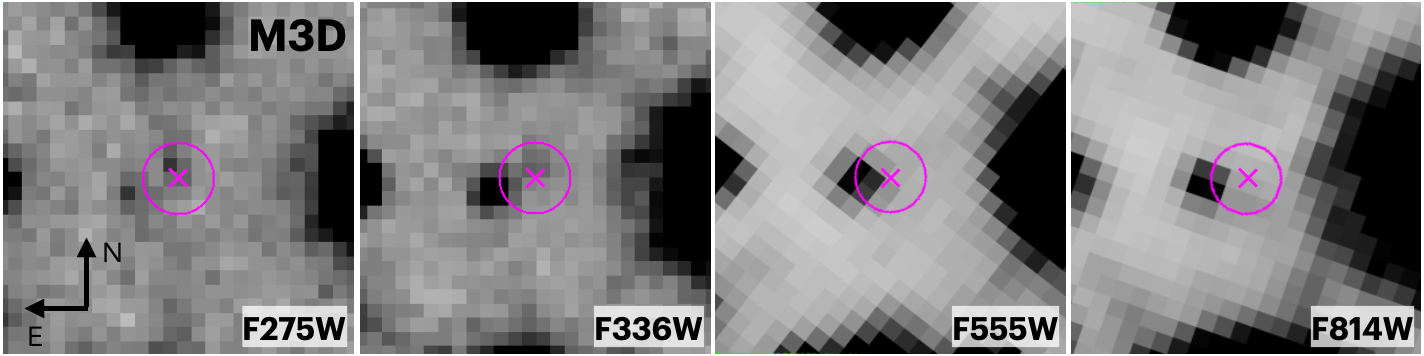}
    \caption{Same as Fig.~\ref{chartM3B} but for M3D.}
    \label{chartM3D}
\end{figure*}

Using our photometric catalog along with the updated, more precise position, we were able to identify the real companion: a blue object lying along the WD cooling sequence in the ($m_{\rm F275W}$, $m_{\rm F275W} - m_{\rm F336W}$) CMD, which is significantly fainter ($m_{\rm F275W} \sim 24.1$ mag) than the optical counterpart found for M3B (see Fig.~\ref{CMD} and \ref{chartM3D}). This object is different from the candidate companion identified by \citet{Cadelano2019}, which is situated approximately $20\arcsec$ away from our object. For this source, photometric measurements were obtained only in the F275W and F336W filters, with no evidence of variability detected in either band.

\begin{figure*}
    \centering
    \includegraphics[width=\hsize]{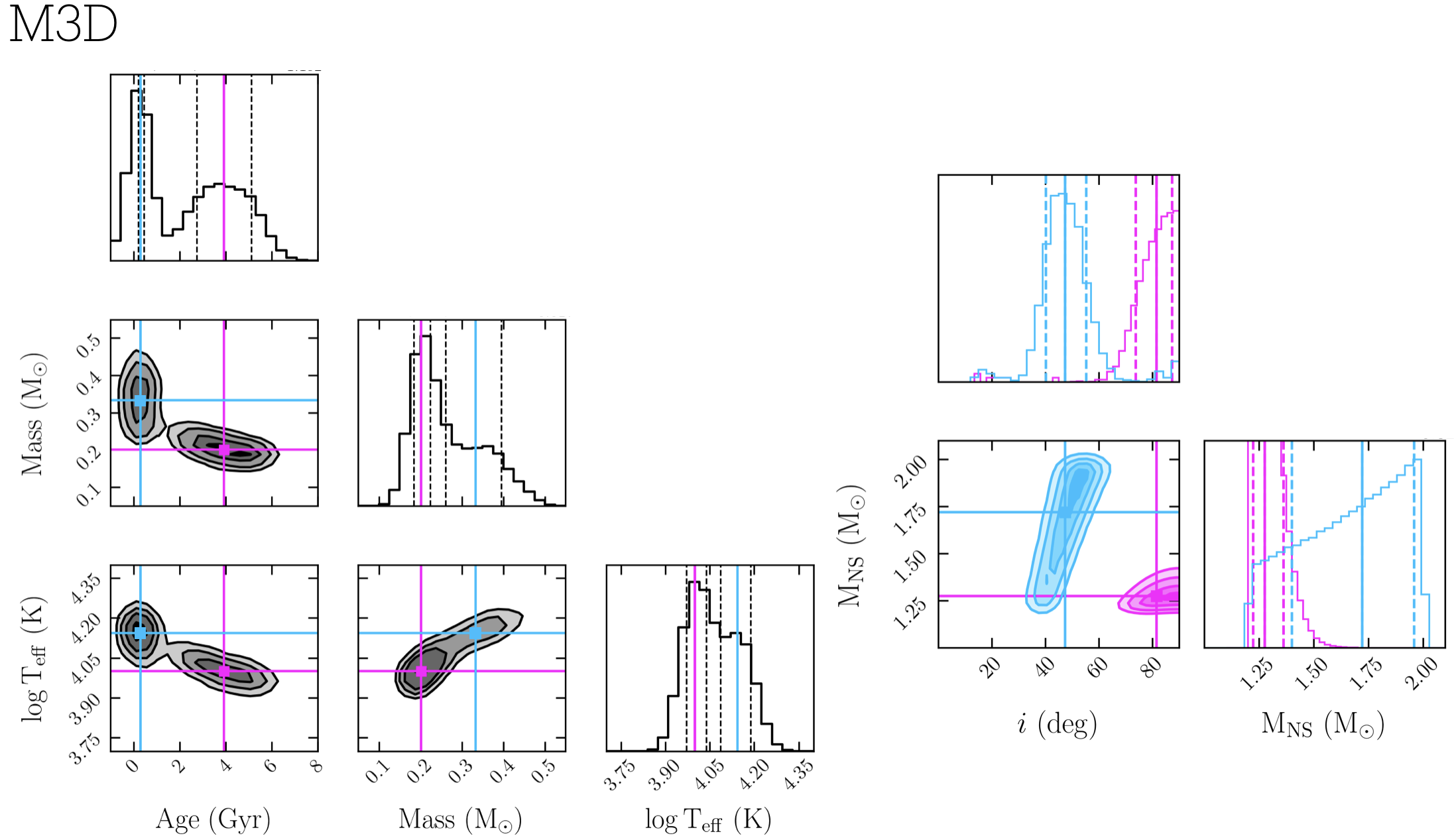}
    \caption{Left-hand corner plots: Same as Fig.~\ref{solutionM3B} but for M3D. Here both the young and old solutions are shown, and they are marked with cyan and magenta lines, respectively. Right-hand corner plot: Constraints on the NS mass and orbital inclination angle of M3D, with posterior distributions shown in cyan for the young solution and magenta for the old solution. The derived values for the companion mass, cooling age, effective temperature, NS mass and orbital inclination for the two solutions are listed in Table~\ref{m3d_m3f_solutions}.}
    \label{solutionM3D}
\end{figure*}

The left-hand panel of Fig.~\ref{solutionM3D} shows the corner plot displaying the 1D and 2D marginalised likelihood distributions for the companion star’s parameters, derived from comparison with theoretical models.
A close inspection of the 1D and 2D likelihood distributions for the cooling age reveals a pronounced, narrow peak around 0.3 Gyr, followed by another peak at $\sim4$ Gyr, similar to the case of M13F in \citet{Cadelano2020}.
The bimodality in the cooling age distribution arises from a dichotomy in the cooling timescales due to the occurrence of diffusion-induced hydrogen-shell flashes in the envelopes of proto-WDs with masses $\mathrm{M} \gtrsim 0.20 \,\Msun$ \citep{Althaus2001a,Althaus2001b,Istrate2014,Istrate2016}. 

In particular, WDs having thinner hydrogen layers are able to cool much faster than those with thicker ones. The models adopted here predict that hydrogen-shell flashes occur only for WDs with masses above approximately $0.24\,\Msun$ \citep[see][]{Cadelano2020}.
Accordingly, the first peak corresponds to masses exceeding $0.24\,\Msun$, which undergo flashes and cool rapidly. In contrast, the secondary peak is linked to lower-mass WDs that avoid flashes and cool more slowly through stable hydrogen burning. 
In this regime, even small photometric uncertainties can make an observation consistent with both pre- and post-flash models. Hence, the bimodal age posterior arises because of the large uncertainties in the companion's measured magnitudes (see Fig.~\ref{CMD}).
When selecting only the young peak of the age likelihood distribution ($M_{COM}>0.24\,\Msun$),  the corresponding companion mass is $M_{COM}= 0.33^{+0.06}_{-0.07}\,\Msun$, the effective temperature $\log{T_{eff}}= 4.14^{+0.05}_{-0.06}$ and $age= 0.31^{+0.15}_{-0.11}\,Gyr$, while the older solution ($M_{COM}\leq0.24\,\Msun$)
is associated with a significantly lower mass companion $M_{COM}= 0.20^{+0.02}_{-0.02}\,\Msun$, with an effective temperature $\log{T_{eff} = 4.00^{+0.04}_{-0.03}}$ and $age= 3.93^{+1.19}_{-1.18}\,Gyr$.
Due to the large photometric uncertainties, the range of possible solutions for M3D span almost the entire range of masses of the adopted model grid. Consequently, it is worth stressing here that the uncertainties on the derived mass, age, and temperature are mostly driven by the prior limits of the model grid rather than by the observational constraints.

Unfortunately, the PSR spin-down age generally cannot provide a more accurate constraint on the system's age, as it is known to be highly unreliable \citep{Tauris2012}. This is due to its dependence on the PSR’s intrinsic spin-down rate, which is difficult to determine for MSPs in GCs because of contamination from acceleration within the cluster potential \citep{Prager2017}. In this case, the situation is further complicated by the lack of a spin-down age measurement for M3D.

The moderate eccentricity ($e \sim 0.075$) and long orbital period ($P_{\rm orb} \sim 129$ d) of M3D suggest that the system has undergone significant dynamical perturbations. These properties might be useful to understand whether M3D is an old system hosting a low-mass WD companion or a relatively younger binary with a more massive WD. Indeed, according to the WD mass - orbital period correlation found by \citet{Tauris1999} for binary radio pulsars, M3D’s long orbital period would favour a system with a more massive WD, hence supporting the young scenario.
Moreover, as discussed by \citet{Hessels2007} and more recently by \citet{Li2024}, binaries with orbital periods between $\sim$10 and 1000 days are often efficiently formed through exchange interactions between an isolated NS and a primordial binary. Such systems tend to survive preferentially in GCs with relatively low stellar densities, like M3 \citep{Hut1992, Sigurdsson1993}. 
In this framework, M3D likely originated from such an exchange encounter, after which the subsequent mass-transfer phase circularized the orbit, as expected for most GC MSPs. A secondary exchange occurring after recycling would typically produce a system with much higher eccentricity and a more massive companion, which is not the case for M3D. Instead, its current mild eccentricity may be explained by cumulative perturbations caused by fly-by encounters occurring after the exchange event, as expected in a relatively low-density cluster like M3.

In this context we can apply equation (5) of \citet{Rasio1995}, which estimates the age of a system given its eccentricity, for $\mathrm{e}\gtrsim0.01$. We assume a central 1D velocity dispersion of $8.2\,\mathrm{km\,s^{-1}}$ and a central mass density of $\log \rho_c = 3.78\,\Msun\,\mathrm{pc^{-3}}$, both taken from the 2023 version of the \citet{Baumgardt2018} GC catalog. We estimate the stellar number density by assuming an average stellar mass of $\sim0.3\,\Msun$. This value was obtained by using the synthetic stellar population provided by the Padova CMD 3.8 web interface\footnote{\url{https://stev.oapd.inaf.it/cgi-bin/cmd_3.8}}, assuming an age of 12.5 Gyr \citep{Dotter2010} and a metallicity of $\mathrm{[Fe/H]}=-1.5$ \citep{Harris2010}. 
This approach does not account for the effects of mass segregation within the cluster. As a result, the derived average mass likely underestimates the true value in the cluster core, where M3D is located. 
The resulting number density of stars is $\mathrm{n}=2.01\times10^{4}\,\mathrm{pc^{-3}}$ and the age of M3D computed using the equation (5) of \citet{Rasio1995} is $\mathrm{t_{e=0.075}}=1.2$ Gyr. It is important to note that this estimate refers to the lifetime of the binary system. Consequently, this value should be interpreted as an upper limit for the age of the WD, favouring the interpretation that the companion of M3D is younger than the age suggested by the second old peak in the probability distribution in Fig.~\ref{solutionM3D}, in agreement with what expected from the WD mass - orbital period relation by \citet{Tauris1999}. However, although the estimated timescale offers useful insight, current uncertainties regarding the formation and evolutionary history of M3D still prevent a definitive determination of its age. 

For both solutions corresponding to the two distinct peaks of the likelihood distributions shown in Fig.~\ref{solutionM3D}, we derived the NS mass and the binary inclination angle by following the same analysis
described in Sect.~\ref{M3B}. When using  $M_{COM}= 0.33^{+0.06}_{-0.07}\,\Msun$, we  find $M_{NS}= 1.72^{+0.24}_{-0.32}\,\Msun$ and $i = 47.33^{+7.91}_{-7.13}\,deg$. On the other side, in the case of a very low-mass and old companion the NS is less massive than in the previous case, with  $M_{NS}= 1.28^{+0.09}_{-0.05}\,\Msun$ corresponding to an edge-on orbit with $i = 81.43^{+5.89}_{-7.88}\,deg$. The posterior distributions or both the young and old solutions are shown in blue and magenta, respectively, in the right-hand panel of Fig.~\ref{solutionM3D}. All values derived for the physical properties of M3D are summarized in Table~\ref{m3d_m3f_solutions}, while the results obtained assuming a bimodal gaussian prior for $\mathrm{M_{NS}}$ are described in Appendix~\ref{appendix1}.

\subsubsection{M3F}\label{M3F}
\begin{figure*}
    \centering
    \includegraphics[width=0.9\hsize]{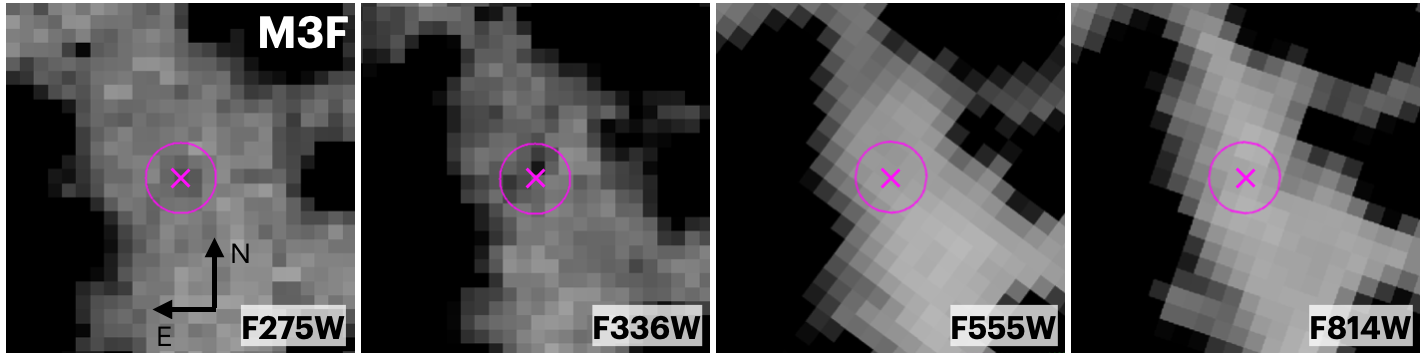}
    \caption{Same as Fig.~\ref{chartM3B} but for M3F.}
    \label{chartM3F}
\end{figure*}

The timing solution derived by \citet{Li2024} for M3F indicates the presence of a companion star with a minimum mass of $0.17\,\Msun$, in a 3-day orbit around the PSR. Searching for the candidate counterpart in our catalog, we identified a very faint star ($m_{\rm F275W} \sim 24.7$ mag), lying along the WD cooling sequence in the ($m_{\rm F275W}$, $m_{\rm F275W} - m_{\rm F336W}$) CMD (marked by the green square in Fig.~\ref{CMD}). This source is detected exclusively in the F275W and F336W filters, and it is shown in the finding charts of Fig.~\ref{chartM3F}. No sign of photometric variability was detected in the available filters.

\begin{figure*}
    \centering
    \includegraphics[width=\hsize]{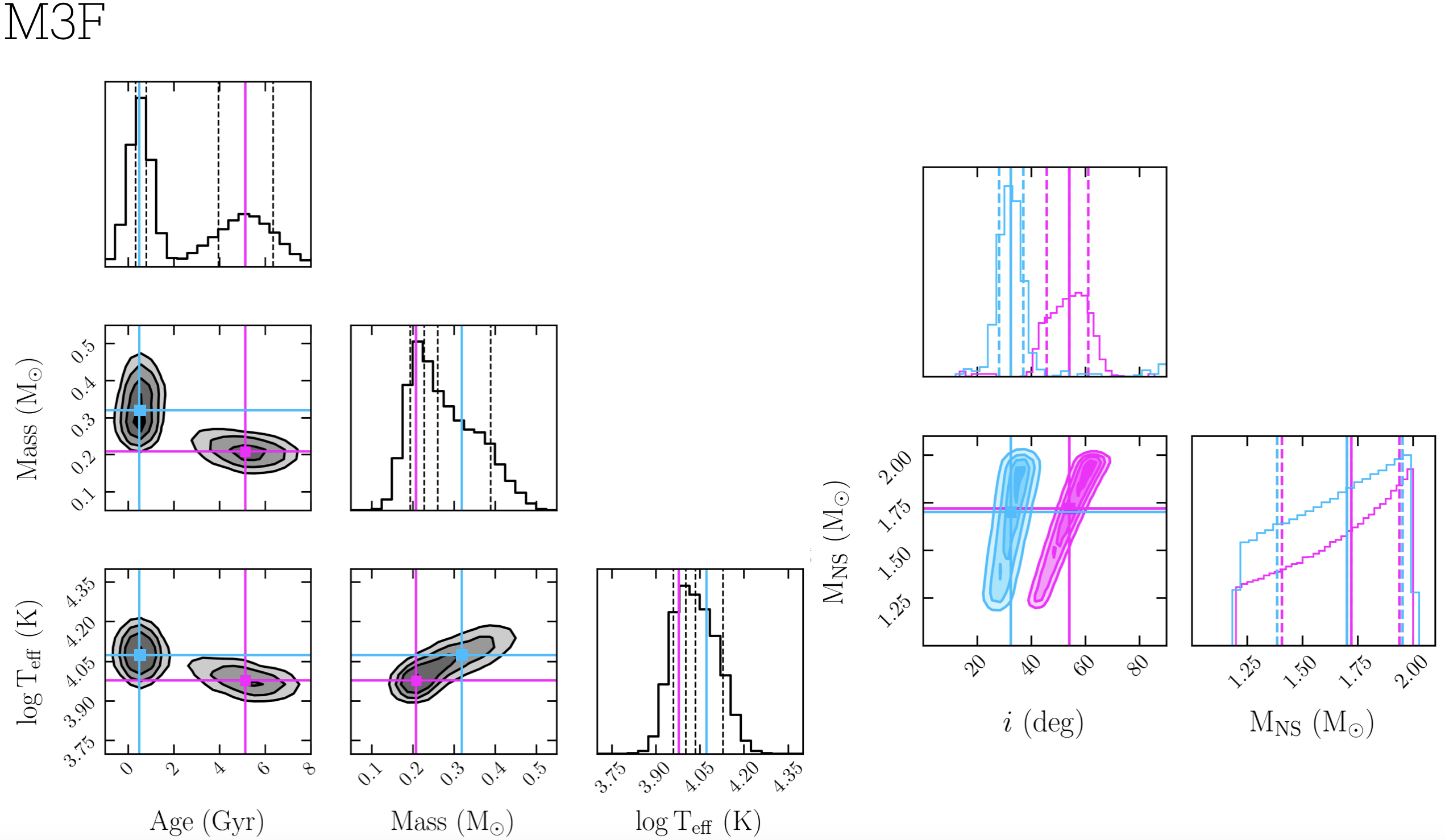}
    \caption{Same as for Fig~\ref{solutionM3D}, but for M3F}
    \label{solutionM3F}
\end{figure*}

We constrained the physical properties of the companion star through comparison with the updated He WD cooling tracks, and, as illustrated by the likelihood distributions of age, $\log{\mathrm{T_{eff}}}$, and $\mathrm{M_{COM}}$ shown on the left side of Fig.~\ref{solutionM3F}, we found a scenario similar to that of M3D. In particular, the 1D histogram of the cooling age shows a clear bimodal distribution, with a narrow peak around 6.2 Gyr, and a secondary smaller peak at $\sim0.5$ Gyr. As already explained in Sect.~\ref{M3D}, this bimodality has a clear physical origin, linked to the onset of diffusion-induced hydrogen-shell flashes in proto-WD around $\mathrm{M} \sim 0.20\,\Msun$. However, the significant photometric uncertainties associated with this faint star affect the mass estimates, broadening the age distribution and preventing the distinction between the two possible age solutions. In this case, selecting only the young peak of the likelihood distribution in Fig.~\ref{solutionM3F} ($M_{COM}>0.24\,\Msun$),  we find a companion with mass $M_{COM}= 0.32^{+0.07}_{-0.06}\,\Msun$, effective temperature $\log{T_{eff}}= 4.07^{+0.06}_{-0.04}$ and $age= 0.51^{+0.29}_{-0.18}\,Gyr$. The older solution ($M_{COM}\leq0.24\,\Msun$)
is instead associated with a significantly lower mass companion $M_{COM}= 0.21^{+0.02}_{-0.02}\,\Msun$, with an effective temperature $\log{T_{eff} = 3.98^{+0.02}_{-0.02}}$ and $age= 5.15^{+1.20}_{-1.19}\,Gyr$.  According to the \citet{Tauris1999} WD mass - orbital period relation, the $\sim 3$ days orbital period of M3F would favour a system with less massive WD, supporting the old scenario.

Similarly to M3D, the large photometric uncertainties associated to M3F result in poorly constrained parameters, with the likelihood distributions mostly reflecting the model boundaries.

We performed the same MCMC analysis as described in Sect.~\ref{M3B}, to simultaneously estimate the NS mass and orbital inclination. For the old solution with a companion mass of $M_{COM} = 0.21^{+0.02}_{-0.02}\,\Msun$ this yielded $M_{NS} = 1.72^{+0.22}_{-0.31}\,\Msun$ and $i = 54.08^{+7.06}_{-8.35}$ degrees. While the younger solution, corresponding to $M_{COM}= 0.32^{+0.07}_{-0.06}\,\Msun$ resulted in a less inclined system, with $i = 32.46^{+4.61}_{-4.34}$ degrees and a slightly less massive NS, of mass $M_{NS} = 1.70^{+0.25}_{-0.32}\,\Msun$.

The posterior distributions corresponding to the two distinct solutions are presented in the right-hand panel of Fig.~\ref{solutionM3F}, while the derived physical parameters for M3F are summarized in Table~\ref{m3d_m3f_solutions}. As for M3D, it is clear that the significant photometric uncertainties impacting the companion mass estimate are preventing our ability to tightly constrain both the orbital inclination and the NS mass, in contrast to the well-determined solution obtained for M3B. Appendix~\ref{appendix1} shows the results obtained for $\mathrm{M_{NS}}$ and $i$ under the assumption of a bimodal gaussian prior for the NS mass.

\subsection{Non detections}\label{nondetections}
\subsubsection{M3A}\label{M3A}
\begin{figure*}
    \centering
    \includegraphics[width=0.9\hsize]{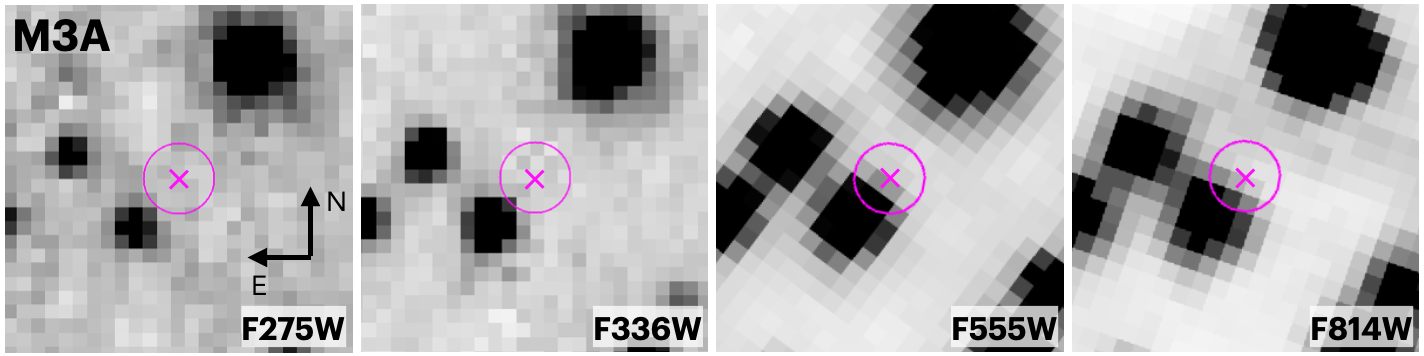}
    \caption{Same as Fig.~\ref{chartM3B} but for M3A.}
    \label{chartM3A}
\end{figure*}

According to \citet{Li2024}, this system is a BW with a minimum companion mass of $0.01\,\Msun$ and an orbital period of 0.14 days. Using our catalog, we searched for potential companions within the defined search radius but found no stellar source at the PSR's position. As shown in the finding chart in Fig.~\ref{chartM3A}, the nearest object is a main sequence star located $\sim0.\!\arcsec2$ away. 

Given the excellent agreement between the radio positions and the astrometry of our catalog, we conclude that this star is unlikely to be the true companion of the MSP. It is more plausible that the actual optical counterpart is either too faint to be detected or obscured by the nearby bright star.  

Therefore, we inspected the PSF-subtracted images obtained after running \texttt{ALLSTAR}, specifically focusing on those taken near the PSR’s inferior conjunction. In fact, BW companions are known for their pronounced optical variability, with light curves typically displaying a distinct minimum and maximum driven by the heating of the companion’s side directly exposed to the PSR’s flux. This variability generally peaks at the PSR's inferior conjunction and reaches a minimum at the superior conjunction \citep{Pallanca2014,Cadelano2015}. However, we did not detect any indication of the true counterpart in the residual images. 

At this point we aimed to set an upper limit on the companion's mass by determining an upper magnitude limit achievable at the PSR position. Since radio timing identifies this system as a BW, we expect the companion to be extremely faint, with most of its emission predominantly in the red filters. Consequently, we estimated a magnitude upper limit from a PSF-subtracted F814W image from HST proposal ID 10775, taken near inferior conjunction. We placed a circular aperture with a radius of $0.5\arcsec$ centered on the radio position of M3A and measured the local number counts.
The 3$\sigma$ upper limit on the total counts within the aperture was computed from the mean and standard deviation of the pixel values inside the aperture. This value was normalized by the exposure time and converted into an instrumental magnitude, which was then calibrated using the filter’s zero point. The resulting 3$\sigma$ upper limit on the calibrated magnitude is $m_\mathrm{{F814W,up}} = 26.06\pm0.02$ mag. This represents the maximum brightness of any undetected optical source at the MSP position.

To constrain the mass of a potential companion, we compared this photometric upper limit with theoretical stellar models. Specifically, we adopted a \textsc{parsec} isochrone \citep{Bressan2012} with an age of 12.5 Gyr \citep{Dotter2010} and a metallicity of $\mathrm{[Fe/H]} = -1.5$ \citep{Harris2010}, consistent with the properties of the cluster. The isochrone was transformed into the observational plane assuming a color excess of $\rm{E(B-V)}=0.01$, a distance of $10.08$ kpc \citep{Baumgardt2021}, and an extinction coefficient of $A_{\mathrm{F814W}}/A_\mathrm{V} = 0.59696$ \citep{Cardelli1989, O'Donnell1994}. By comparing the calibrated limiting magnitude at the PSR position with the interpolated isochrone, we derived an upper limit on the companion mass of $\mathrm{M_{COM,up}} = 0.108 \pm 0.001\,\Msun$. 

\subsubsection{M3E}\label{M3E}
MSP M3E is in an orbit of 7.1 days with a minimum companion mass of $0.2\,\Msun$. The timing solution for this system was presented for the first time by \citet{Li2024}. Starting from the radio position, we found no stellar source within $0.\!\arcsec1$ in any filter, with the exception of the images from proposal IDs 10775 (PI: Sarajedini) and 16298 (PI: Libralato). 

\begin{figure}
    \centering
    \includegraphics[width=0.9\hsize]{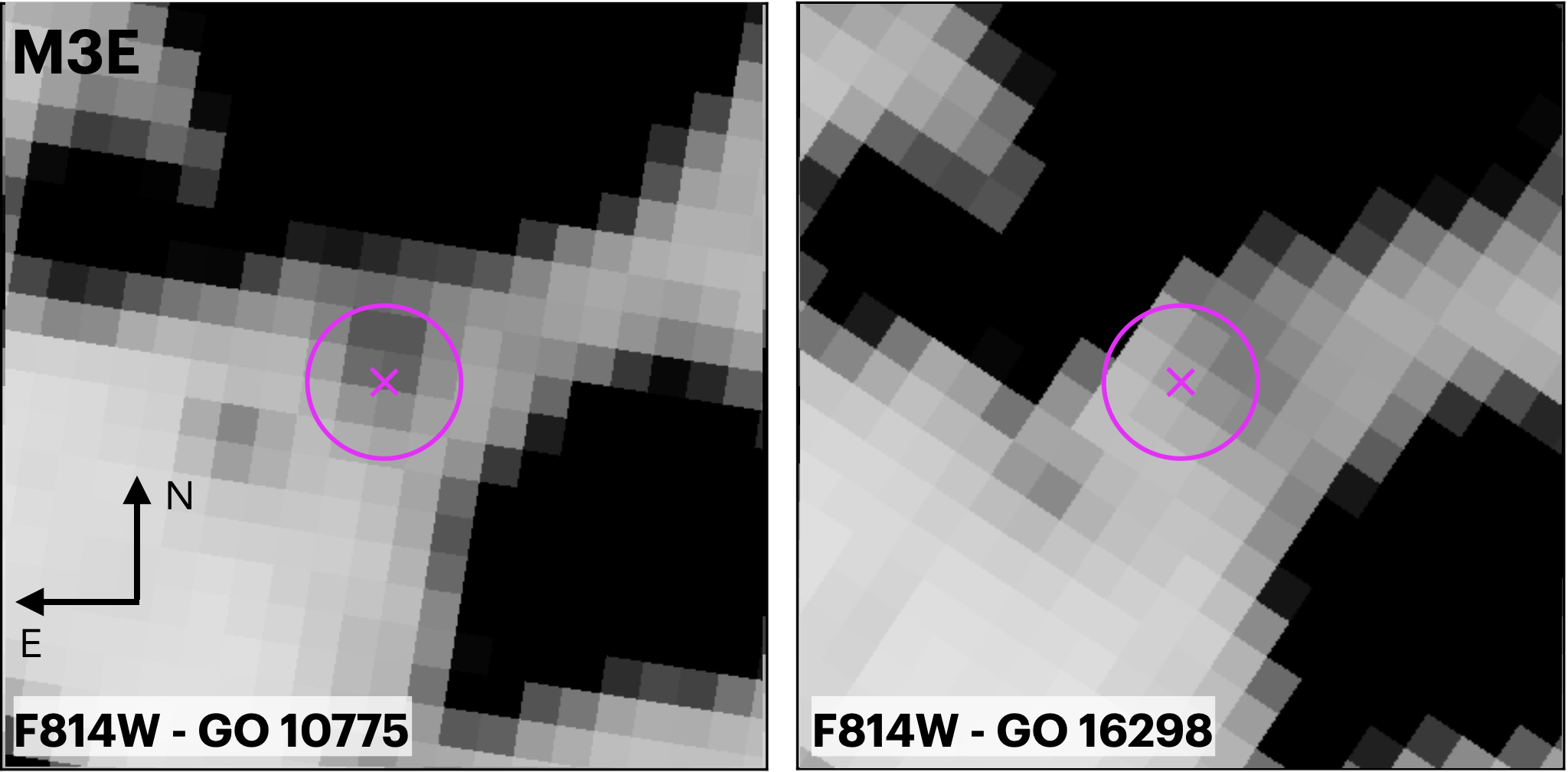}
    \caption{Finding charts for the MSP M3E. The panels show the \texttt{$\_$drc} images in the ACS/WFC F814W filter from HST proposal ID 10775 (left) and ID 16298 (right). Each panel spans $1\arcsec\times1\arcsec$ centered on the PSR position.}
    \label{chartM3E}
\end{figure}

In both datasets, a source is located very close to the radio position, at a distance of $\sim0.\!\arcsec03$. This object is detected only in the F814W band images, suggesting it is a red source, predominantly emitting at longer wavelengths. However, as shown in the finding chart in Fig.~\ref{chartM3E}, its position clearly shifts between the two epochs, separated by approximately 15 years.

\begin{figure}
    \centering
    \includegraphics[width=0.9\hsize]{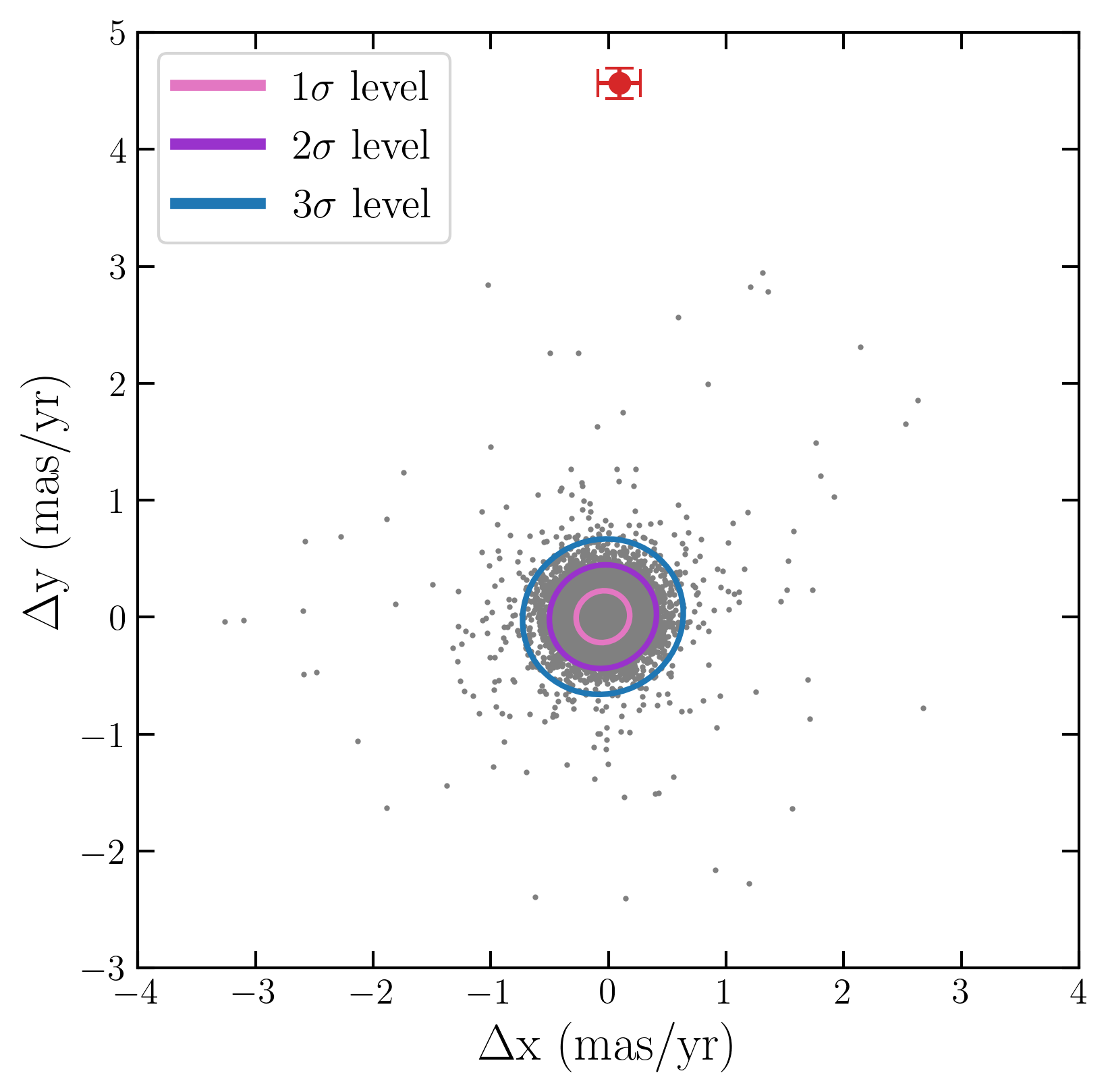}
    \caption{Vector point diagram showing the proper motion displacements in x and y for all stars common to the HST datasets from proposals 10775 and 16298 (grey points). The red dot marks the position and uncertainty of the object at the M3E radio position. The $1\sigma$, $2\sigma$, and $3\sigma$ contours, representing the modeled cluster PM distribution, are shown for reference. The distribution was modeled with a two-dimensional Gaussian Mixture Model (GMM) using a single component, implemented via the XDGMM Python framework \citep{holoien2017}.}
    \label{propermotions}
\end{figure}

As anticipated in Sect.~\ref{analysis}, these two datasets were used to measure relative stellar PMs. This allowed us to determine the PM of the object located at the position of M3E. The resulting vector point diagram is shown in Fig.\ref{propermotions}, where grey points correspond to stars detected in both epochs, and the red dot indicates the motion of the object at the M3E radio position, including its associated uncertainty. Its measured motion ($\Delta\mathrm{x} = 0.091 \pm 0.060$ mas/yr; $\Delta\mathrm{y} = 4.567 \pm 0.044$ mas/yr) lies well outside the cluster's proper motion locus, strongly suggesting that this object is clearly not a cluster member. 

According to \citet{Li2024}, the proper motions and distances of the PSRs in M3 have not been individually measured. However, their values are expected to be consistent with those of the cluster itself, as suggested by the remarkably similar dispersion measures of the PSRs. Consequently, if M3E is a cluster member, the detected source near its radio position cannot be its true optical counterpart. Instead, its significant relative proper motion and red color strongly indicate that it is a foreground object unrelated to the cluster.

As in the case of M3A, we attempted to estimate an upper limit on the magnitude of the undetected companion. Given the uncertain nature of the object, we translated this constraint into an upper limit on the companion’s mass by comparing the magnitude limit with both a \textsc{parsec} isochrone, assuming the companion is a non-degenerate star, and the binary evolution models from \citet{Cadelano2020}, under the alternative hypothesis that the companion is a WD.
To do so, we used  WFC3/UVIS F275W, F336W and F438W PSF-subtracted images from HST program GO-12605, adopting the same approach described in Sect.~\ref{M3A}.

Comparing the magnitude upper limit obtained for the three filters ($m_\mathrm{{F275W,up}} = 25.55\pm 0.01$ mag, $m_\mathrm{{F336W,up}} = 24.26\pm 0.01$ mag and $m_\mathrm{{F438W,up}} = 24.12\pm 0.01$ mag) with the \textsc{parsec} isochrone described in Sect.~\ref{M3A}, we inferred a corresponding upper limit on the companion mass of $\mathrm{M_{COM,up}} = 0.484 \pm 0.001\,\Msun$.
Under the WD hypothesis, we selected from the \citet{Cadelano2020} model grid only those models that predict magnitudes fainter than the observed upper limits in all three WFC3 filters.  Each consistent model was assigned equal probability, and the posterior distribution for the companion mass was derived using weighted percentiles, adopting the 84th percentile as the mass limit.  In the WD scenario, because more massive WDs are intrinsically fainter at a given cooling age, an undetected companion must be fainter, and therefore more massive, than our detection limit, resulting in a lower limit on its mass. Therefore we obtained $M_{COM,lower} = 0.390 \pm 0.001\,\Msun$.

\section{Conclusions}\label{Conclusions}
In this work, we carried out a detailed search for the optical counterparts to five binary MSPs located in the GC  M3 (NGC~5272). Our analysis builds upon the recent high-precision, phase-connected timing solutions presented by \citet{Li2024}, which provide accurate PSR positions and orbital parameters. We built an extensive HST photometric dataset, spanning from near-ultraviolet (F275W) to optical I-band (F814W), to carry out an in-depth multi-band investigation of each MSP. For each system, we searched for optical sources within a $0.\!\arcsec1$ radius of the timing position, a conservative threshold that accounts for HST astrometric uncertainties. Candidate counterparts were evaluated based on: (i) positional coincidence with the PSR, (ii) their location in CMDs, and (iii) comparison with updated binary evolution models \citep[see][]{Cadelano2020}.

For M3A and M3E, no optical counterparts were identified, likely due to crowding by nearby bright stars or intrinsic faintness of the companions. To set upper limits on the companion mass, we estimated a magnitude upper limit at the PSR positions using PSF-subtracted images. 
For M3A, a BW system, we derived an upper companion mass limit of $\mathrm{M_{COM,up}} = 0.108 \pm 0.001\,\Msun$ by comparing the limiting magnitude to a suitable \textsc{parsec} isochrone.
In the case of M3E, a red source detected in F814W images at the radio position was ruled out as a foreground contaminant based on PM measurements. Given the uncertain nature of the companion, we used both \textsc{parsec} isochrones and the binary evolution models of \citet{Cadelano2020} to translate the magnitude limits into companion mass limits. We found $M_{COM,lower} = 0.390 \pm 0.001\,\Msun$ under the WD hypothesis and $\mathrm{M_{COM,up}} = 0.484 \pm 0.001\,\Msun$ in the case of a non-degenerate companion.

On the other side, we identified and characterized the optical counterparts to three MSPs: M3B, M3D, and M3F.  While the optical companion to M3B was previously studied by \citet{Cadelano2019}, we reanalyzed this system using updated photometry and improved evolutionary models to refine its parameters. For M3D, the optical counterpart we identify is significantly offset (by $\sim20\arcsec$) from the candidate proposed by \citet{Cadelano2019}. This discrepancy originates from the large positional uncertainties in both R.A. and Dec. associated with the timing solution of M3D from \citet{Hessels2007}, which was adopted in the earlier study. Finally, the timing solution for M3F was only recently obtained by \citet{Li2024}, making this work the first attempt to identify its companion star.

In all three cases, the counterpart lies precisely at the radio position, demonstrating excellent agreement between optical and radio astrometry, and they exhibit photometric properties consistent with low-mass He WDs in NS binary systems. 
Based on model comparison, we confirm that M3B is a low-mass He WD with a relatively short cooling age of $0.98^{+0.18}_{-0.19}$ Gyr and a mass of $\mathrm{M_{COM}} =0.18^{+0.01}_{-0.01}\,\Msun$. All the properties inferred for M3B are in agreement with the results presented by \citet{Cadelano2019}.

For M3D and M3F, we report the first identification of their optical companions. In both cases, the counterparts are very faint blue stars whose magnitudes and colors are fully compatible with those of cooling He WDs. Both objects exhibit a bimodal cooling age distribution, with prominent peaks at ages $\lesssim 1$ Gyr and $\sim 4-5$ Gyr. This bimodality arises from a dichotomy in cooling timescales caused by diffusion-induced hydrogen-shell flashes. However, the significant photometric uncertainties associated with these two faint companions propagate into the physical parameter estimates, broadening the age distributions and hindering a clear distinction between the old and young solutions. When selecting only the young peak of the age likelihood distribution of M3D, the corresponding companion mass is $M_{COM}= 0.33^{+0.06}_{-0.07}\,\Msun$, while the older solution
is associated with a significantly lower mass companion $M_{COM}= 0.20^{+0.02}_{-0.02}\,\Msun$. A similar result is obtained for M3F, with companion masses of 
$M_{COM}= 0.32^{+0.07}_{-0.06}\,\Msun$ and $M_{COM}= 0.21^{+0.02}_{-0.02}\,\Msun$, for the young and old solutions, respectively.

It is important to note here that there is a substantial difference in the meaning of the physical parameters derived for the three systems and more specifically of their uncertainties. For M3B, the tight constraints from both magnitude and color measurements locate the companion within a well-defined region of the model grid. Consequently, the inferred mass, cooling age, and effective temperature represent robust physical determinations with meaningful uncertainties. In contrast, the substantially larger photometric errors for M3D and M3F make these companions virtually consistent with nearly the full range of masses in the adopted model grid, as clearly illustrated in Fig.~\ref{CMD}. The resulting posterior distributions for mass, age, and temperature and relative uncertaintes, derived in the two separate young and old solutions, are therefore mostly driven by the prior boundaries of our grid rather than by the constraining power of the data. This caveat should be kept in mind when interpreting the fitted values for these two systems.

Although our analysis does not fully break the degeneracy between the orbital inclination and the NS mass, we are able to reduce the range of plausible solutions by combining the derived companion mass with the binary's orbital parameters. We employ an affine-invariant MCMC ensemble sampler, adopting a flat prior on the orbital inclination and a flat prior for $\mathrm{M_{NS}}$, within the range of $1.2-2.0\,\Msun$, consistent with the NS mass distribution found by \citet{Abbott2023}. For M3B we found an almost edge-on configuration ($i = 88.84^{+1.16}_{-6.77}$ deg) with a NS mass of $\mathrm{M_{NS}}=1.22^{+0.02}_{-0.01}\,\Msun$. For M3D and M3F, we inferred these two parameters separately for the young and old solutions, finding NS masses around $\sim 1.7\,\Msun$  and low to moderately inclined systems with $i$ between $\sim30$ deg and $\sim50$ deg. The only exception is the solution found for M3D, under the old hypothesis, showing a low-mass NS ($M_{NS}= 1.28^{+0.09}_{-0.05}\,\Msun$) and an edge-on inclination ($i = 81.43^{+5.89}_{-7.88}\,deg$).

To resolve the degeneracy between NS mass and inclination angle, independent measurements of the companion mass, and ideally the orbital inclination, are required. A possible approach would be to determine the mass ratio of the binary components through spectroscopic measurements, but the companion's very low luminosity places this beyond the reach of current optical telescopes. At the same time, the radio detection of the Shapiro time delay could yield precise masses for both the MSP and its companion star \citep{Corongiu2012,Fonseca2021,Corongiu2023}, especially for highly inclined systems like M3B, however, precise timing and dense radio observations are required.

This study highlights the effectiveness of combining high-precision radio timing with deep, multi-band HST imaging. The synergy between radio and optical observations, supported by comparisons with updated evolutionary models, enables the identification and characterization of MSP companions even under challenging observational conditions.
However, as clearly demonstrated by the cases of M3D and M3F, future efforts to reach fainter magnitudes with significantly improved photometric precision are essential. Only by reducing uncertainties we can achieve a secure and detailed characterization of the faintest counterparts, as photometric errors play a crucial role in limiting our ability to constrain the nature of the entire system. This goal will become increasingly achievable with the advent of next-generation facilities such as the Extremely Large Telescope, whose MORFEO adaptive optics system and MICADO camera will deliver the sensitivity and angular resolution needed to overcome current limitations in the most crowded and faint observational regimes.

\begin{acknowledgements}
The research activities described in this paper were carried out with contribution of the Next Generation EU funds within the National Recovery and Resilience Plan (PNRR), Mission 4 - Education and Research, Component 2 - From Research to Business (M4C2), Investment Line 3.1 - Strengthening and creation of Research Infrastructures, Project IR0000034 – ``STILES - Strengthening the Italian Leadership in ELT and SKA''.
GE and ED also acknowledge financial support from the INAF Data analysis Research Grant (PI E. Dalessandro) of the “Bando Astrofisica Fondamentale 2024”. CP acknowledges support from the INAF Large Grant 2022 “GCjewels” (P.I. Andrea Possenti) approved with the Presidential Decree 30/2022. We thank the referee for their comments and suggestions, which have helped us improve the quality and clarity of the manuscript. G.E. is grateful to A. Della Croce for the useful discussions.

\end{acknowledgements}
\bibliographystyle{aa} 
\bibliography{biblio} 
\begin{appendix}
\onecolumn
\section{Bimodal gaussian prior on the NS mass}\label{appendix1}
\begin{figure*}[h!]
    \centering
    \includegraphics[width=\hsize]{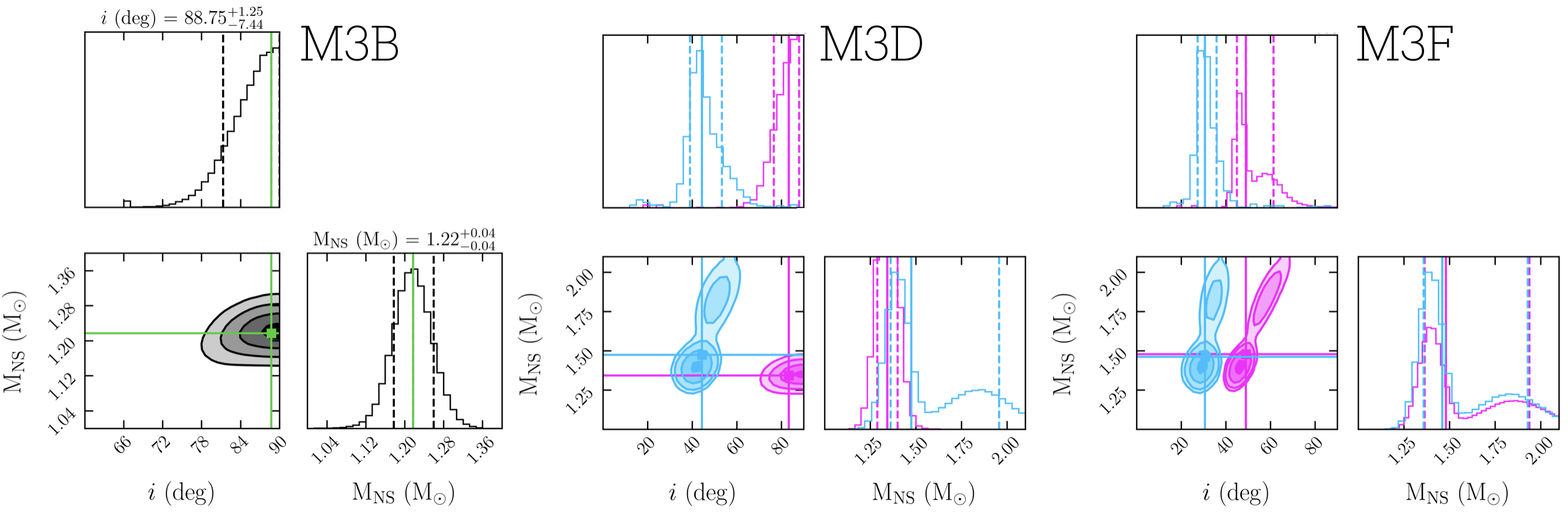}
    \caption{Constraints on the NS mass and the orbital inclination angle of M3B, M3D and M3F, from left to right, assuming a flat prior for the inclination angle and a bimodal gaussian prior for the NS mass, following \citet{Antoniadis2016}. The one-dimensional histograms represent the marginalised probability distributions for these two parameters, with the solid blue and dashed black lines indicating the best-fit values and their associated uncertainties. The bottom-left panel shows the joint two-dimensional posterior probability distribution, with contours representing the 1$\sigma$, 2$\sigma$, and 3$\sigma$ confidence levels. For M3D and M3F, the posterior distributions shown in cyan for the young solution and magenta for the old solution. The derived values for the companion mass, cooling age, effective temperature, NS mass and orbital inclination for the two solutions are listed in Table~\ref{m3d_m3f_solutions_appendix}.}
    \label{appendix_fig}
\end{figure*}

Figure~\ref{appendix_fig} shows the posterior distributions for the NS mass and the orbital inclination angle for  for the three systems with identified optical counterparts: M3B, M3D, and M3F (from left to right). These results were obtained assuming a flat prior on the cosine of the inclination angle and a bimodal Gaussian prior on the NS mass, peaking at $\sim1.4\,\Msun$ and $1.8\,\Msun$ and uniformly sampled between $1$ and $2.5\,\Msun$ \citep{Antoniadis2016}. This choice of prior differs from the analysis presented in the main text, where a flat prior between $1.2$ and $2\,\Msun$ was adopted, following the NS mass distribution inferred by \citet{Abbott2023} from gravitational wave observations (see Sect.~\ref{M3B}).

On one side, the flat-prior approach removes any prior assumptions about the expected value of $\mathrm{M_{NS}}$, allowing the data to explore a broader, physically plausible mass range, and yielding results that reflect a deliberately non-informative perspective. On the other hand, adopting a bimodal Gaussian prior enables us to test whether the observed properties of each system are consistent with a canonical NS mass distribution, while also providing an estimate of the orbital inclination under this assumption, restricting the parameter space to values consistent with the typical mass range observed in recycled MSPs. The results for the three systems are summarized in Table~\ref{m3d_m3f_solutions_appendix}.

Under the latter assumption, M3B shows a very narrow posterior on both the inclination angle and the NS mass, with the best-fit values consistent with the ones inferred under the assumption of a flat prior for $\mathrm{M_{NS}}$, yielding a NS mass of $1.22^{+0.04}_{-0.04}\,\Msun$ and a moderately high inclination angle ($88.75^{+1.25}_{-7.44}$ deg). This confirms that the system is observed at high inclination angle and hosts a NS with a mass consistent with the typical values measured for recycled NSs in compact binary systems.

For M3D and M3F, we show the posterior distributions of $\mathrm{M_{NS}}$ and $i$ corresponding to the two companion masses derived from the young and old solutions in cyan and magenta, respectively, consistently with Figs.~\ref{solutionM3D} and ~\ref{solutionM3F} (see Sections~\ref{M3D} and \ref{M3F}).

For both systems, the posterior distributions are much broader with respect to the case of M3B, with the NS mass posterior distribution that still reflects the shape of the adopted bimodal prior. The only exception is the posterior distribution of M3D in the old companion scenario, constraining a nearly edge-on system with $i=83.27^{+4.67}_{-6.70}$ deg and a NS mass of $1.34^{+0.06}_{-0.05}\,\Msun$, similarly to what is obtained assuming a flat prior on the NS mass. On the other side, the old solution of M3F corresponds to a relatively inclined system ($48.97^{+12.43}_{-4.00}$ deg) and a NS with mass $\mathrm{M_{NS}}=1.48^{+0.46}_{-0.12}\,\Msun$.
Regarding the young solutions, the best-fit values indicate relatively low-inclination systems ( $44.44^{+8.86}_{-5.40}$ deg and $30.70^{+5.14}_{-3.31}$ deg, for M3D and M3F respectively), hosting less massive NS ($1.47^{+0.48}_{-0.11}\,\Msun$ and $1.46^{+0.47}_{-0.10}\,\Msun$, respectively) compared to the results obtained assuming a flat prior on $\mathrm{M_{NS}}$ (see Sect.~\ref{M3D} and \ref{M3F}).

These results once more confirm that the degeneracy in the physical parameters observed for both M3D and M3F significantly restricts our ability to constrain $\mathrm{M_{NS}}$ and the system’s orbital geometry.
Hence, achieving higher photometric precision, and thereby obtaining more accurate measurements of the companion star’s properties, is crucial for improving the determination of the system’s properties.

\renewcommand{\arraystretch}{1.3}
\begin{table*}[h!]
\setlength{\tabcolsep}{6pt}
\caption{\label{m3d_m3f_solutions_appendix}Inferred NS mass and orbital inclination for M3B, M3D and M3F, assuming a bimodal gaussian distribution for the prior on $\mathrm{M_{NS}}$. For the M3D and M3F systems, we show both the young and old solutions. }
\centering
\begin{tabular}{lccccc}
\hline\hline
& M3B & \multicolumn{2}{c}{M3D} & \multicolumn{2}{c}{M3F} \\
\cmidrule(lr){3-4} \cmidrule(lr){5-6}
Parameter & &Young & Old & Young & Old \\
\hline
$\mathrm{M_{NS}}\,(\Msun)$ & $1.22^{+0.04}_{-0.04}$&$1.47^{+0.48}_{-0.11}$ & $1.34^{+0.06}_{-0.05}$ & $1.46^{+0.47}_{-0.10}$ & $1.48^{+0.46}_{-0.12}$ \\
$i$ (deg) & $88.75^{+1.25}_{-7.44}$&$44.44^{+8.86}_{-5.40}$ & $83.27^{+4.67}_{-6.70}$ & $30.70^{+5.14}_{-3.31}$ & $48.97^{+12.43}_{-4.00}$ \\
\hline
\end{tabular}
\end{table*}
\end{appendix}
\end{document}